\documentclass[12pt]{article}

\usepackage{scicite}
\usepackage{xcolor}
\usepackage{pdfpages}
\usepackage{url}
\usepackage[version=3]{mhchem} 
\usepackage{amsmath,amssymb,lscape}
\usepackage{alltt}
\usepackage[T1]{fontenc}
\usepackage{graphicx}
\usepackage[]{lmodern}
\usepackage{stmaryrd}
\usepackage[sc]{mathpazo}
\usepackage{times}
\usepackage{float}
\usepackage{url}
\usepackage{hyperref}
\usepackage{caption}
\usepackage{indentfirst}
\usepackage{pdfpages}
\usepackage[labelfont=bf, labelsep=period]{caption}

\newenvironment{sciabstract}{%
\begin{quote} \bf}
{\end{quote}}

\title{Closed-Loop Design of Proton Donors for Lithium-Mediated Ammonia Synthesis with Interpretable Models and Molecular Machine Learning}

\author
{Dilip Krishnamurthy,$^{1\dagger}$ Nikifar Lazouski,$^{2\dagger}$ Michal L. Gala, $^{2}$\\ Karthish Manthiram$^{2\ast}$ and Venkatasubramanian Viswanathan$^{1\ast}$\\
\\
\normalsize{$^{1}$Department of Mechanical Engineering,}\\ \normalsize{Carnegie Mellon University, Pittsburgh, PA, USA}\\
\normalsize{$^{2}$Department of Chemical Engineering,}\\ \normalsize{Massachusetts Institute of Technology, Cambridge, MA, USA}\\
\\
\normalsize{$\dagger$Equally Contributing Authors; $^\ast$E-mails: karthish@mit.edu, venkvis@cmu.edu}
}

\date{}

\begin{document} 


\baselineskip24pt


\maketitle


\begin{sciabstract}

In this work, we experimentally determined the efficacy of several classes of proton donors for lithium-mediated electrochemical nitrogen reduction in a tetrahydrofuran-based electrolyte, an attractive alternative method for producing ammonia. We then built an interpretable data-driven classification model, which identified solvatochromic Kamlet-Taft parameters as important for distinguishing between active and inactive proton donors. After curating a dataset for the Kamlet-Taft parameters, we trained a deep learning model to predict the Kamlet-Taft parameters. The combination of classification model and deep learning model provides a predictive mapping from a given proton donor to the ability to produce ammonia. We demonstrate that this combination of classification model with deep learning is superior to a purely mechanistic or a data-driven approach in accuracy and experimental data efficiency.

\end{sciabstract}

\subsection*{Introduction}

Ammonia is an industrial chemical that is used to produce a variety of nitrogen-containing compounds, such as fertilizers, pharmaceuticals, and polymers \cite{fao2017world,erisman2008century}. In addition to being a useful synthetic molecule, ammonia (NH$_3$) is also emerging as an attractive carbon-free energy carrier, as it can be liquefied at moderate pressures (>10 bar) at room temperature \cite{soloveichik2014liquid,Jiao2019}; the volumetric density of liquid ammonia greatly exceeds that of lithium-ion batteries and is competitive with other chemical storage media, such as pressurized and liquid hydrogen \cite{Gur2018}. NH$_3$ is typically produced via the Haber-Bosch process at high temperatures (450-550$^{\circ}$ C) and pressures (up to 200 bar)\cite{giddey2017ammonia}. The process produces up to 1.44\% of the world's carbon dioxide emissions due to its use of fossil fuels as a hydrogen source \cite{giddey2017ammonia,Soloveichik2019} and is economically viable only in large, centralized plants\cite{foster2018catalysts}.\newline 

With dramatically falling renewable energy prices \cite{IRENA2020}, there is an incentive to produce NH$_3$ in a distributed manner using renewable energy. Electrochemical methods have been proposed to produce ammonia in a distributed manner from intermittent power sources with no CO$_2$ emissions and low capital costs \cite{Soloveichik2019}. While a large number of catalyst compositions and electrolyzer configurations have been proposed for electrochemical nitrogen reduction \cite{McPherson2019,Shi2020}, many of them report Faradaic efficiencies and production rates too low for practical utilization. In addition, there are numerous calls for rigorous controls and reproducibility in the electrochemical nitrogen reduction field \cite{Andersen2019,Greenlee2018,MacLaughlin2019,Shipman2017}, suggesting that ammonia is often detected from adventitious sources. \newline

Methods utilizing lithium metal as a mediator report some of the highest Faradaic efficiencies (FEs) and absolute rates, as well as strict and reproducible controls, of proposed electrochemical approaches for NH$_3$ synthesis\cite{tsuneto1994lithium,mcenaney2017ammonia,ma2017reversible,kim2018electrochemical}. In this approach, lithium metal is first produced via electrochemical reduction of lithium ions (Li$^+$), which spontaneously breaks the nitrogen triple bond to produce lithium nitride\cite{greenwood1997chemistry}; this lithium nitride can then react with a proton donor to form ammonia, recovering lithium ions (Fig. \ref{fig1:cycleSetup}a). The approach has been demonstrated to produce ammonia in both batchwise \cite{mcenaney2017ammonia,ma2017reversible,kim2018electrochemical} and continuous systems \cite{tsuneto1994lithium,Lazouski2019,Schwalbe2020,Andersen2019}. \newline

While the proton donor is responsible for protonating lithium nitride, there are reasons to believe that its role goes beyond being a source of hydrogen in ammonia, for instance in activating the lithium nitridation reaction \cite{McFarlane1962,Lazouski2019,tsuneto1994lithium}. A previous theoretical analysis of a general nitrogen reduction reaction demonstrated that the thermodynamic activity of the proton donor is important for selective continuous nitrogen reduction\cite{Singh2019}. A preliminary survey of proton donors demonstrated that the identity of the proton donor has a profound effect on the ammonia yields in the lithium-mediated nitrogen reduction reaction (LM-NRR).\cite{tsuneto1994lithium} A similar effect has been observed for other reduction reactions involving proton donors \cite{Peters2019,Jiang2020,Ren2019}. However, no detailed studies of effect of the identity of the proton donor on LM-NRR have been performed to date. It is essential to identify molecular design rules for the proton donor in order to improve selectivity of the process toward ammonia.\newline

Approaches to discovering material design rules typically involve learning a physics-based functional mapping from material choice to performance through governing equations that represent relevant physical laws and underlying interactions. As these approaches rely on physical laws, they are rationalizable or interpretable; certain approximations or empirical terms pertaining to hard-to-encode interactions may need to be added for model accuracy. On the other hand, with significant increases in computational power, several studies across disciplines have demonstrated the effective use of deep learning models to learn the material-to-performance mapping with increased predictive power albeit with limited interpretability.\cite{meredig2018can,ward2016general,balachandran2018predictions,ward2018machine,ren2018accelerated} The enhancement in predictive power is in part attributed to the ability of deep architecture models to accurately learn physical interactions that are otherwise difficult to explicitly encode with a functional form. \newline



Our work provides a novel approach for designing the electrolyte composition to enable electrochemical ammonia production via a lithium-mediated approach. We develop an interpretable deep-learning-based model to predict the ability of a proton donor, which is a primary electrolyte component, to activate nitrogen reduction. The model forms a part of the closed-loop methodology between experiments and computation to predict good candidates, provide feedback from testing, refine the model, and discover novel proton donors of the electrolyte. Results from human-intuition-driven experimental exploration were used to generate a data-driven interpretable model to identify relevant physical descriptors. Through experimental testing, we found 1-butanol can promote LM-NRR to a greater extent than the conventionally-used ethanol. Through a data-driven approach, we identified most notably that Kamlet-Taft parameters, denoted as $\alpha$ and $\beta$ \cite{kamlet1976solvatochromic,taft1976solvatochromic}, are descriptors of the ability of the proton donor to promote ammonia formation.  The developed closed-loop methodology between experiments and computation is expected to be beneficial to the broader community and can be leveraged to find novel materials for electrochemical routes in the production of other chemicals.

\subsection*{Experimental characterization of proton donors for ammonia production}

\begin{figure}[H]
		\centering
\includegraphics[width=0.75\linewidth]{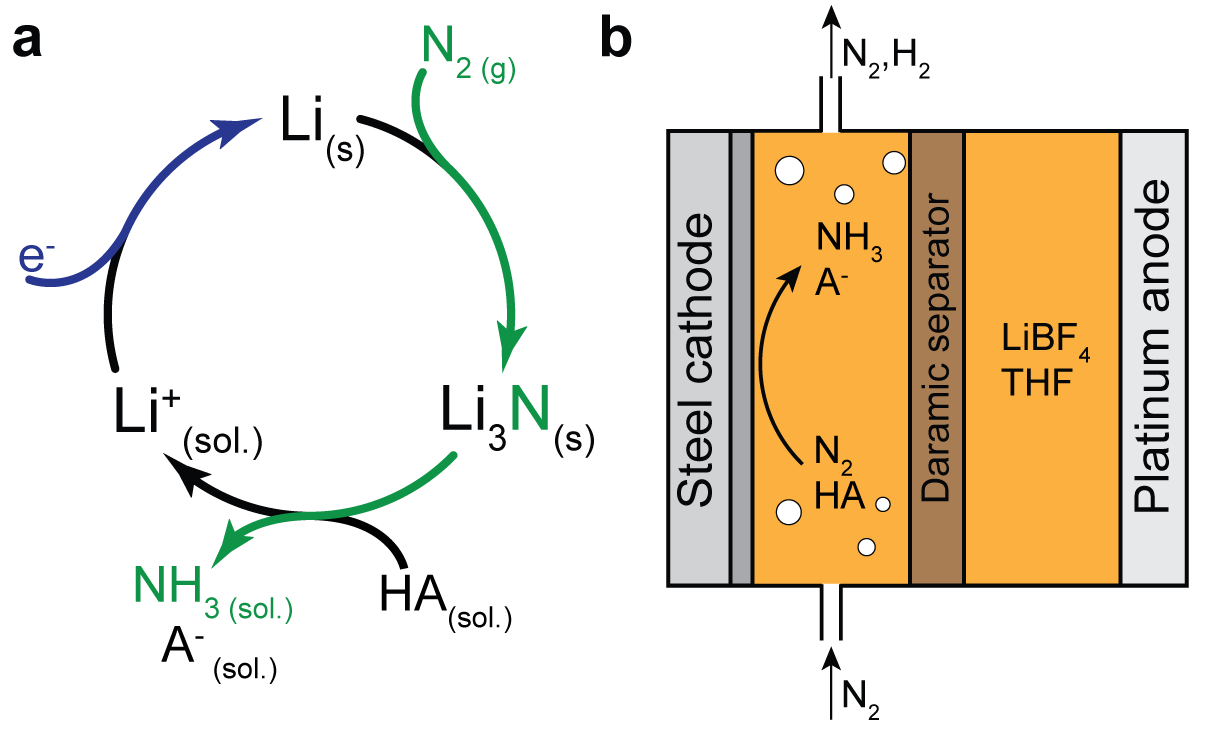}
		\caption{\textbf{Lithium-mediated ammonia production from nitrogen.} \textbf{(a)} The lithium-mediated catalytic cycle, with species flows highlighted. \textbf{(b)} The electrochemical cell setup used for continuous ammonia production and proton donor testing.} 
		\label{fig1:cycleSetup}
\end{figure}

The presence of a proton donor in the electrolyte during LM-NRR is necessary for converting fixed nitrogen in the form of lithium nitride to ammonia (Fig. \ref{fig1:cycleSetup}a). As lithium nitride and its derivatives, imide and amide, are strong bases (p\textit{K}$_{\text{a}}$ = 41 for NH$_3$ \cite{bordwell1981}), it is likely that many compounds irrespective of structure can thermodynamically promote liberation of ammonia from nitride. However, there appears to exist a threshold concentration of proton donor below which nitride ammonia and lithium nitride may not be detected following electrolysis of a lithium-ion-containing solution, particularly when ethanol is used as the proton donor \cite{Lazouski2019,tsuneto1994lithium}. This suggests that the proton donor plays a secondary role in LM-NRR, i.e. that it promotes the reaction to fix the nitrogen, either electrochemically or thermochemically. The ability of of a proton donor to promote nitrogen fixation appears to depend on its structure \cite{tsuneto1994lithium}. \newline

In order to determine whether a given proton donor can promote LM-NRR, a number of candidate compounds were tested at flooded stainless steel electrodes at a range of concentrations in a previously described setup\cite{Lazouski2019}. Briefly, a 1 M LiBF$_4$ in tetrahydrofuran (THF) electrolyte was used in a 2-compartment electrochemical cell with a platinum foil anode, stainless steel foil cathode, and polyporous Daramic separator (Fig. \ref{fig1:cycleSetup}b, Fig. S1). A range of concentrations of proton donor were added to the electrolyte prior to electrolysis. Nitrogen gas was flowed through the the cathode compartment while a constant current was applied across the cell (Fig. \ref{fig1:cycleSetup}b). If the proton donor promotes nitrogen reduction, then ammonia forms and can be detected in the electrolyte via a colorimetric assay (see Supplementary Methods). \newline

We decided that a proton donor is classified as active toward in LM-NRR if the Faradaic efficiency (FE) towards ammonia in at least one operating condition exceeds 0.5\%; if all experiments lead to FEs below 0.5\%, then the proton donor is considered inactive. This threshold was chosen based on the minimum quantifiable FE ($\sim$0.1\%) and the spread in FE typically observed at low production rates ($\sim$0.1\%); a threshold value of 0.5\% increases the likelihood that a given proton source is indeed active for LM-NRR when ammonia is detected and reduces the likelihood that the ammonia signal is spurious or comes from adventitious sources \cite{Du2019,Hu2019,Shipman2017,Dabundo2014}. For a detailed description of experimental procedures, see the Supplementary Methods. \newline

As it is resource-intensive and time-consuming to test a large number of conditions and compounds for activity in LM-NRR, there is a need for determining what factors are important for activity to reduce the number of necessary experiments. As the differences in activity are a function of proton donor structure, several simple hypotheses could be proposed to explain the differences in activity between various classes of compounds. For instance, one could propose that the activity of the proton donor is correlated to its acidity (p\textit{K}\textsubscript{a} value). For highly acidic donors, such as carboxylic acids, the reaction between lithium metal and the proton donor, or even direct electrochemical reduction of the proton donor to hydrogen gas without lithium deposition, may be favored over the nitrogen reduction reaction. Weakly acidic proton donors, on the other hand, may be inert in electrochemical reactions or reactions involving lithium (e.g., the reaction between t-butanol and lithium is slow \cite{Talalaeva1964}), thus not promoting nitrogen reduction significantly. Therefore, an intermediate p\textit{K}$_{\text{a}}$ value could be desired for nitrogen reduction. In light of this, p\textit{K}$_{\text{a}}$ and other potential descriptors were examined for the ability to distinguish between active and inactive proton donors. No significant classification ability was observed for simple descriptors chemical and steric descriptors such as p\textit{K}$_{\text{a}}$ and Bader volume (Fig. S4). As simple descriptors proved to be insufficient for describing the differences in tested proton donors towards nitrogen reduction, we turned to a more rigorous, data-driven approach to find activity correlations from experimental results


\subsection*{Identifying desirable properties of proton donors}

We employed a data-driven approach to determine favorable properties of proton donors towards promoting LM-NRR. Several quantitative properties of proton donors were assessed including measures of solvent strength as acids or bases (acid dissociation constant (p\textit{K}$_{,\text{a}}$), Guttman donor and acceptor numbers \cite{khetan2015trade,khetan2014identifying,gutmann1976solvent}), measures of reactivity (highest occupied molecular orbital (HOMO) and lowest unoccupied molecular orbital (LUMO) levels \cite{mccloskey2012limitations,khetan2017effect,khetan2014identifying}), solvatochromatic Kamlet-Taft (KT) parameters ($\alpha$, $\beta$, $\pi^*$ \cite{crowhurst2006using,wilson1991using}), measures of ionic character and polarizability (Bader charge), and computable measures of steric hindrance and diffusivity (Bader volume)\cite{pande2019descriptors,garcia2012importance}. Using these properties, we built and trained a range of models (linear and non-linear supervised learning models, regression models, decision trees) to predict the experimentally observed binary activity classification for the initial set of compounds. We curated a training dataset composed of all the aforementioned properties curated from existing literature\cite{marcus1993properties,meyer1995correlation,Sten} and our density functional theory (DFT) calculations. We then found that a decision tree (Fig. \ref{fig3:classModelPred}a) which takes KT parameters as inputs was associated with high classification ability ($\approx$ 96\% accuracy) in addition to being interpretable. Specifically, KT parameters denoted as $\alpha$ and $\beta$, which quantify solvent hydrogen-bond donating and accepting ability, were identified as descriptors of activity. The decision tree was optimized through cross-validation to balance tree complexity and misclassification error of the model (see Supporting Information for details).   \newline


The obtained decision tree (Fig. \ref{fig3:classModelPred}a) identifies a simple criterion for above-threshold activity towards electrochemical ammonia production: $\alpha>\alpha_t=0.78$ and $\beta>\beta_t=0.59$. The identified classification can be rationalized based on the fact that promising proton donors should exhibit both high proton donating ($\alpha>\alpha_t$) and accepting ability ($\beta>\beta_t$). Such a criterion can be rationalized as the key nitrogen fixation reaction (6Li + N$_2$ $\rightarrow$ 3Li$_3$N) involves formation of undercoordinated lithium ions (Li$^+$), the closest chemical analogue to a proton, during formation of lithium nitride; these ions can be stabilized by the basicity of the proton donor ($\beta$), thus accelerating nitrogen fixation. The need for a threshold solvent acidity ($\alpha$) can be rationalized by the fact that the nitrogen must be protonated to ultimately produce ammonia; stabilization of deprotonated forms of nitrogen during reduction may accelerate the fixation reaction. Alternatively, proton donating character may be necessary for promoting the formation of defect sites in the lithium metal, which may be necessary for formation of lithium nitride \cite{McFarlane1962}. An inherent proton donating-accepting trade-off emerges in the $\alpha-\beta$ space (Fig. \ref{fig3:classModelPred}b), where only a small fraction of candidates strike a balance above identified threshold values. \newline

A majority of compounds identified to be promising for ammonia production from the first set of experimentally tested candidates are recovered (Table S3), which indicates the robustness of the developed classification model. Several additional candidates with experimentally known KT parameter values were then tested as to more accurately determine $\alpha_T$ and $\beta_T$ (Fig. \ref{fig3:classModelPred}, Tables S2 and S4). This closed-loop refinement of the interpretable model (Fig. \ref{fig5:closedLoop}) was performed thrice after initial experiments, which decreased the uncertainty in fitted $\alpha_T$ and $\beta_T$. The limited number of proton donors exceeding these threshold KT parameter values (Fig. \ref{fig3:classModelPred}b, Fig. S5) highlights that identifying novel candidates is challenging due to the narrow diversity of chemical structures that occur within these thresholds for $\alpha$ and $\beta$. 


\subsection*{Deep-learning framework for prediction of KT parameters}

Given that KT parameters were identified as key descriptors of the ability of proton donors to promote LM-NRR and that they are experimentally known only for a limited set of compounds, we developed a deep learning model to predict these parameters for arbitrary compounds in order to assess the activity of a wide range of candidate donors. While other approaches to predict KT parameters have been proposed,\cite{waghorne2020study,sherwood2019method} their associated computational cost is significantly high for exploring a large chemical space. Our model was trained on a curated dataset of compounds for which experimentally measured values for $\alpha$ and $\beta$ are reported in the literature\cite{marcus1993properties,meyer1995correlation,Sten}; the dataset size was $n=222$ compounds (low-data regime), thereby requiring careful and robust model training using an ensemble of models. Using an ensemble of models, i.e. a population of independently trained models with varied initial starting configurations, allowed us to quantify the uncertainty of predictions for novel compounds and families of compounds.\newline

We employed a deep learning-based model as implemented in the DeepChem package \cite{Ramsundaretal2019} to predict the KT parameters since deep learning models have proven useful especially in the low-data regime\cite{moret2020generative}. In addition, the mapping from molecular features to activity is likely high-dimensional due to the complex underlying physics. The model was trained on molecular features from candidates using the simplified molecular-input line-entry system (SMILES) representation. Out of several molecular featurization approaches based on SMILES representations, the weave featurization coupled with a weave model (deep neural network)\cite{kearnes2016molecular} led to the most accurate (training and testing RMSEs of $\approx$0.01) and generalizable predictions, i.e. ones with low cross-validation error, of $\alpha$ and $\beta$ (Fig. \ref{fig4:deepLearningPred}). The weave featurization\cite{wu2018moleculenet} encodes both local chemical environment and connectivity of atoms in a molecule. A detailed description of the featurization scheme, neural network architecture, and training routines can be found in the Supplementary Information. \newline



The deep learning model (material-descriptor relationship) coupled with the classification model (descriptor-activity relationship) was used to predict the activity of tested and novel proton donors (Fig. \ref{fig3:classModelPred}). In order to evaluate the robustness of predictions associated with various proton donors and to determine promising candidates to experimentally test, for each candidate we computed the $c$-value (confidence value)\cite{houchins2017quantifying} from an ensemble of deep learning models. The $c$-value for a given material, $c_M$, is computed as the fraction of ensemble models that predict the candidate to exhibit desirable performance, which in this context can be defined as 

\begin{equation}
c_{M} = \frac{\textit{\# of models with desirable parameters}}{\textit{\# of models in ensemble}} = \frac{1}{N_{ens}} \sum_{n=1}^{N_{ens}}\Theta (\alpha_{pred,n}^M-\alpha_{T})\Theta(\beta_{pred,n}^M-\beta_{T}).   
\end{equation}

where, $N_{ens}$ is the number of models in the ensemble, $\Theta$  is the Heaviside function, $\alpha_{pred,n}^M$ and $\beta_{pred,n}^M$ are the predicted values of KT parameters from the $n^{\mathrm{th}}$ model in the ensemble, and $\alpha_T$ and $\beta_T$ represent the threshold values identified by the classification model. The approach involving an ensemble of models allows us to identify candidates for which there is disagreement between individual models, which indicates additional training data is necessary for higher certainty. \newline

The solvatochromic parameters $\alpha$ and $\beta$ were predicted for 1,000,000 compounds from the PubChem database. We observed that a large fraction of the compounds have predicted KT parameter values that lie outside the region described by the decision tree obtained from experiments. Only $\sim$0.54\% of the 1 million compounds have $c$-values exceeding 0.5 and only $\sim$0.19\% have $c$-values exceeding 0.7, suggesting that compounds which the models predict to be active with a high degree of confidence are rare. Linear aliphatic alcohols, which were experimentally determined to be active in LM-NRR, are recovered through the models with high $c$-values. However, the vast majority of candidates with high $c$-values are biological compounds with both hydrogen bond donating (hydroxyl) and accepting (amine) groups, hence their large $\alpha$ and $\beta$. These candidates could not be tested for activity in LM-NRR as they contained nitrogen and were not readily commercially available. \newline

As many candidates with high $c$-values could not be tested due to the aforementioned reasons, the goal of further experiments was to learn the descriptor-activity relationship or the delineating surface between active and inactive candidates with greater accuracy. It is worth highlighting that after every batch of performed experiments, the experimental activity was used to augment the input data to the classification model in a closed-loop fashion to more accurately learn and update the descriptor-activity relationship (Fig. \ref{fig5:closedLoop}). \newline

\subsection*{Experimental validation of models}
In order to assess and improve the predictive capability of the interpretable decision tree and deep learning models, we selected a number of the candidates from various regions of the $\alpha$-$\beta$ space for experimental validation (Fig. \ref{fig6:expTesting}). Several proton donors with known KT parameters above the identified threshold values were tested as a result of the decision tree model. Novel proton donors for which experimentally measured KT parameters were not known, but were estimated to be promising using the deep-learning model, were also tested; a total of seven candidates were selected for further experimental testing. Two proton donors were predicted to be inactive with a large degree of confidence ($c\approx0$): formic acid and ethyl acetate. Five proton donors were predicted to be active with various degrees of confidence ($c=0.17-0.67$): 2-ethyl-1-butanol, triethylene glycol, 2,2-dimethyl-1,3-propanediol, 4-methoxybutan-1-ol, and 1,4-cyclohexanedimethanol. \newline

Formic acid and ethyl acetate were found to not promote nitrogen reduction, as predicted by the decision-tree and deep-learning models, due to their extreme $\alpha$ and $\beta$ values (Fig. \ref{fig6:expTesting}). Of the five potentially active candidates tested, two were found to promote LM-NRR: 2-ethyl-1-butanol and 2,2-dimethyl-1,3-propanediol produced ammonia with FEs of 3.62\% and 0.84\%, respectively (Fig. \ref{fig6:expTesting}). The ether-group-containing compounds were found to be inactive, even despite relatively large $c$-values, especially in the case of triethylene glycol. While both of the ether-group-containing compounds have above-threshold $\alpha$ and $\beta$ values, they have higher $\alpha$ and lower $\beta$ than the two active aliphatic alcohols (Table S3). This may suggest the relative importance of having a high $\beta$ or balanced $\alpha$ and $\beta$ values for proton donors that are active in LM-NRR.  \newline

\subsection*{Discussion}

Differences between experimentally measured and predicted proton donor activity could have several origins. One reason could be changes in the proton donors' structures under the very reductive conditions of the experiment (in the presence of lithium metal). For instance, halogenated compounds could easily decompose to form lithium halide salts on the electrode, preventing further electrochemical reactions, as well as changing the structure of the proton donor significantly. Analyzing the effect of these decomposition pathways is outside the scope of this study. Alternatively, the model correlating the activity with bulk KT parameters may not be all-encompassing; there may be additional descriptors that may need to be incorporated into the model to fully capture the differences between proton donors. The decision tree from data-driven modeling maximized classification performance while striking the right complexity-performance trade-off based on decision tree optimization through cross-validation; the inclusion of additional nodes to the model was identified to lead to overfitting of the model and an associated lower information criterion. Additional experimental data using predicted candidates could prove to be useful and necessitate the inclusion of additional parameters or altering the values of existing parameters to accurately predict experimental results (Fig. \ref{fig3:classModelPred}). \newline

We highlight that the two-model approach involving the material-descriptor mapping (deep-architecture model) in conjunction with the descriptor-activity mapping (shallow-learning model) is a novel paradigm in material discovery. The shallow-learning model allows the interpretation of identified descriptors. In the current work, the ability of the solvatochromic parameters to describe the activity towards ammonia production is rationalized based on mechanistic hypotheses regarding the key nitrogen fixation reaction. Alongside, the deep-learning model has the ability to capture the potentially non-linear mapping between a given material and the descriptors. 

A purely deep-learning approach to predict experimental activity directly from tested compounds would be limited to a few tens of experimental training data. A key advantage in the current approach is its ability to learn the mapping on hundreds of relevant experimentally-derived training data pertaining to solvatochromic parameters, the importance of which was shown via the interpretable model. On the other hand, a purely mechanistic approach (shallow model) would enable activity prediction only on a few hundred materials for which solvatochromic parameters are known. The developed methodology allows interpretability while enabling predictions on a vast set of materials with the ensemble of models as a way to calibrate the associated confidence.


\subsection*{Conclusions}
In the present work, we determined the effect of chemical structure of proton donors on lithium-mediated nitrogen reduction by testing a number of families of proton donors for activity. After experimentally testing a number of families of proton donors, some structure-activity trends were observed. Through the testing, 1-butanol was discovered as the most effective proton donor for LM-NRR. After failing to explain observed structure-activity trends with simple parameterization models, a rigorous data-driven driven approach was used to identify descriptors of activity towards ammonia production. Solvatochromic Kamlet-Taft parameters $\alpha$ and $\beta$ were found to best describe proton donors' ability to promote nitrogen reduction, leading to an interpretable classification model involving the two parameters. The fact that these solvatochromic parameters emerge as the descriptors can be rationalized based on the mechanistic hypothesis that the solvent's hydrogen bond donating (captured by $\alpha$) and accepting (captured by $\beta$) ability are important in the key reaction of simultaneous lithium ion stabilization and protonation of nitrogen by the proton donor. \newline

A deep learning based model (material-descriptor mapping) was developed on molecular features to predict $\alpha$ and $\beta$ values for any given compound, and was used in conjunction with the classification model (descriptor-activity mapping) to perform a vast search for promising candidates from about 1M compounds. Through a closed-loop approach, candidates were proposed for experimental testing with the primary goal of best learning the delineating surface (in $\alpha-\beta$ space) between active and inactive candidates. The loop between computation and experiments was closed by data augmentation after every batch of experimental testing. After the initial experimentation phase four loops were carried out (Fig. \ref{fig5:closedLoop}) with batches of experiments performed each time towards learning the material-activity relationship. The closed-loop approach between experiments and theory has enabled an increase in the fraction of tested active candidates from 30\% during the initial exploration to 65\% during the combined effort. In the process, several novel active proton donors were discovered, demonstrating the robustness and power of the coupled experimental-data driven approach to studying complex systems. \newline

We believe that the approach presented in this work can be utilized for studying and improving a range of chemical and catalytic systems. While the approach identified descriptors based on strong correlations with desired outputs, sufficiently strong correlations help build hypotheses for mechanistic understanding of complex chemical processes. By using a limited set of experimental data, we are able to determine experimental parameters than can predict and affect future experiments (Fig. \ref{fig3:classModelPred}). An automated method for predicting the values of relevant parameters allows for rapid identification of potential leads. By testing novel leads, the model and understanding of the process can be significantly improved when compared to a conventional, intuition-driven experimental approach.

\clearpage
\bibliography{scibib}

\begin{thebibliography}{10}

\bibitem{fao2017world}
F.~FAO, {\it et~al.\/}, {\it Food and Agriculture Organization of the United
  Nations.\/}  (2017).

\bibitem{erisman2008century}
J.~W. Erisman, M.~A. Sutton, J.~Galloway, Z.~Klimont, W.~Winiwarter, {\it Nat.
  Geosci.\/} {\bf 1}, 636 (2008).

\bibitem{soloveichik2014liquid}
G.~L. Soloveichik, {\it Beilstein J. Nanotechnol.\/} {\bf 5}, 1399 (2014).

\bibitem{Jiao2019}
F.~Jiao, B.~Xu, {\it Adv. Mater.\/} {\bf 31}, 1805173 (2019).

\bibitem{Gur2018}
T.~M. G{\"{u}}r, {\it Energy Environ. Sci.\/} {\bf 11}, 2696 (2018).

\bibitem{giddey2017ammonia}
S.~Giddey, S.~Badwal, C.~Munnings, M.~Dolan, {\it ACS Sustain. Chem. Eng.\/}
  {\bf 5}, 10231 (2017).

\bibitem{Soloveichik2019}
G.~Soloveichik, {\it Nat. Catal.\/} {\bf 2}, 377 (2019).

\bibitem{foster2018catalysts}
S.~L. Foster, {\it et~al.\/}, {\it Nat. Catal.\/} {\bf 1}, 490 (2018).

\bibitem{IRENA2020}
IRENA, {Renewable Power Generation Costs in 2019}, {\it Tech. rep.\/} (2020).

\bibitem{McPherson2019}
I.~J. McPherson, T.~Sudmeier, J.~Fellowes, S.~C.~E. Tsang, {\it Dalton
  Trans.\/} {\bf 48}, 1562 (2019).

\bibitem{Shi2020}
L.~Shi, Y.~Yin, S.~Wang, H.~Sun  (2020).

\bibitem{Andersen2019}
S.~Z. Andersen, {\it et~al.\/}, {\it Nature\/} {\bf 570}, 504 (2019).

\bibitem{Greenlee2018}
L.~F. Greenlee, J.~N. Renner, S.~L. Foster, {\it ACS Catal.\/} {\bf 8}, 7820
  (2018).

\bibitem{MacLaughlin2019}
C.~MacLaughlin, {\it ACS Energy Lett.\/} pp. 1432--1436 (2019).

\bibitem{Shipman2017}
M.~A. Shipman, M.~D. Symes, {\it Electrochim. Acta\/} {\bf 258}, 618 (2017).

\bibitem{tsuneto1994lithium}
A.~Tsuneto, A.~Kudo, T.~Sakata, {\it J. Electroanal. Chem.\/} {\bf 367}, 183
  (1994).

\bibitem{mcenaney2017ammonia}
J.~M. McEnaney, {\it et~al.\/}, {\it Energy Environ. Sci.\/} {\bf 10}, 1621
  (2017).

\bibitem{ma2017reversible}
J.-L. Ma, D.~Bao, M.-M. Shi, J.-M. Yan, X.-B. Zhang, {\it Chem\/} {\bf 2}, 525
  (2017).

\bibitem{kim2018electrochemical}
K.~Kim, {\it et~al.\/}, {\it ChemSusChem\/} {\bf 11}, 120 (2018).

\bibitem{greenwood1997chemistry}
N.~Greenwood, A.~Earnshaw, {\it Chemistry of the Elements 2nd Edition\/}
  (Butterworth-Heinemann, 1997).

\bibitem{Lazouski2019}
N.~Lazouski, Z.~J. Schiffer, K.~Williams, K.~Manthiram, {\it Joule\/} {\bf 3},
  1127 (2019).

\bibitem{Schwalbe2020}
J.~A. Schwalbe, {\it et~al.\/}, {\it ChemElectroChem\/} p. celc.201902124
  (2020).

\bibitem{McFarlane1962}
E.~F. McFarlane, F.~C. Tompkins, {\it Trans. Faraday Soc.\/} {\bf 58}, 997
  (1962).

\bibitem{Singh2019}
A.~R. Singh, {\it et~al.\/}, {\it ACS Catal.\/} {\bf 9}, 8316 (2019).

\bibitem{Peters2019}
B.~K. Peters, {\it et~al.\/}, {\it Science\/} {\bf 363}, 838 (2019).

\bibitem{Jiang2020}
C.~Jiang, A.~W. Nichols, J.~F. Walzer, C.~W. Machan, {\it Inorg. Chem.\/} {\bf
  59}, 1883 (2020). PMID: 31935070.

\bibitem{Ren2019}
S.~Ren, {\it et~al.\/}, {\it Science\/} {\bf 365}, 367 (2019).

\bibitem{meredig2018can}
B.~Meredig, {\it et~al.\/}, {\it Mol. Syst. Des. Eng.\/} {\bf 3}, 819 (2018).

\bibitem{ward2016general}
L.~Ward, A.~Agrawal, A.~Choudhary, C.~Wolverton, {\it npj Comput. Mater.\/}
  {\bf 2}, 1 (2016).

\bibitem{balachandran2018predictions}
P.~V. Balachandran, {\it et~al.\/}, {\it Phys. Rev. Mater.\/} {\bf 2}, 043802
  (2018).

\bibitem{ward2018machine}
L.~Ward, {\it et~al.\/}, {\it Acta Mater.\/} {\bf 159}, 102 (2018).

\bibitem{ren2018accelerated}
F.~Ren, {\it et~al.\/}, {\it Sci. Adv.\/} {\bf 4}, eaaq1566 (2018).

\bibitem{kamlet1976solvatochromic}
M.~J. Kamlet, R.~Taft, {\it J. Am. Chem. Soc.\/} {\bf 98}, 377 (1976).

\bibitem{taft1976solvatochromic}
R.~Taft, M.~J. Kamlet, {\it J. Am. Chem. Soc.\/} {\bf 98}, 2886 (1976).

\bibitem{bordwell1981}
F.~G. Bordwell, G.~E. Drucker, H.~E. Fried, {\it J. Org. Chem.\/} {\bf 46}, 632
  (1981).

\bibitem{Du2019}
H.-L. Du, T.~R. Gengenbach, R.~Hodgetts, D.~R. MacFarlane, A.~N. Simonov, {\it
  ACS Sustain. Chem. Eng.\/} {\bf 7}, 6839 (2019).

\bibitem{Hu2019}
B.~Hu, M.~Hu, L.~Seefeldt, T.~L. Liu, {\it ACS Energy Lett.\/} {\bf 4}, 1053
  (2019).

\bibitem{Dabundo2014}
R.~Dabundo, {\it et~al.\/}, {\it PLoS One\/} {\bf 9}, e110335 (2014).

\bibitem{Talalaeva1964}
T.~V. Talalaeva, G.~V. Tsareva, A.~P. Simonov, K.~A. Kocheshkov, {\it Bulletin
  of the Academy of Sciences, USSR Division of Chemical Science\/} {\bf 13},
  595 (1964).

\bibitem{khetan2015trade}
A.~Khetan, A.~Luntz, V.~Viswanathan, {\it J. Phys. Chem. Lett.\/} {\bf 6}, 1254
  (2015).

\bibitem{khetan2014identifying}
A.~Khetan, H.~Pitsch, V.~Viswanathan, {\it J. Phys. Chem. Lett.\/} {\bf 5},
  1318 (2014).

\bibitem{gutmann1976solvent}
V.~Gutmann, {\it Coord. Chem. Rev.\/} {\bf 18}, 225 (1976).

\bibitem{mccloskey2012limitations}
B.~D. McCloskey, {\it et~al.\/}, {\it J. Phys. Chem. Lett.\/} {\bf 3}, 3043
  (2012).

\bibitem{khetan2017effect}
A.~Khetan, H.~Pitsch, V.~Viswanathan, {\it Phys. Rev. Mater.\/} {\bf 1}, 045401
  (2017).

\bibitem{crowhurst2006using}
L.~Crowhurst, R.~Falcone, N.~L. Lancaster, V.~Llopis-Mestre, T.~Welton, {\it J.
  Org. Chem.\/} {\bf 71}, 8847 (2006).

\bibitem{wilson1991using}
L.~Y. Wilson, G.~R. Famini, {\it J. Med. Chem.\/} {\bf 34}, 1668 (1991).

\bibitem{pande2019descriptors}
V.~Pande, V.~Viswanathan, {\it J. Phys. Chem. Lett.\/} {\bf 10}, 7031 (2019).

\bibitem{garcia2012importance}
M.~Garc{\'\i}a-Mota, {\it et~al.\/}, {\it J. Phys. Chem. C\/} {\bf 116}, 21077
  (2012).

\bibitem{marcus1993properties}
Y.~Marcus, {\it Chem. Soc. Rev.\/} {\bf 22}, 409 (1993).

\bibitem{meyer1995correlation}
P.~Meyer, G.~Maurer, {\it Ind. Eng. Chem. Res.\/} {\bf 34}, 373 (1995).

\bibitem{Sten}
R.~Stenutz, Kamlet-taft solvent parameters,
  \url{http://www.stenutz.eu/chem/solv26.php} (accessed June 2020).

\bibitem{waghorne2020study}
W.~E. Waghorne, {\it J. Solution Chem.\/} pp. 1--20 (2020).

\bibitem{sherwood2019method}
J.~Sherwood, J.~Granelli, C.~R. McElroy, J.~H. Clark, {\it Molecules\/} {\bf
  24}, 2209 (2019).

\bibitem{Ramsundaretal2019}
B.~Ramsundar, {\it et~al.\/}, {\it Deep Learning for the Life Sciences\/}
  (O'Reilly Media, 2019).
  \url{https://www.amazon.com/Deep-Learning-Life-Sciences-Microscopy/dp/1492039837}.

\bibitem{moret2020generative}
M.~Moret, L.~Friedrich, F.~Grisoni, D.~Merk, G.~Schneider, {\it Nat. Mach.
  Intell.\/} {\bf 2}, 171 (2020).

\bibitem{kearnes2016molecular}
S.~Kearnes, K.~McCloskey, M.~Berndl, V.~Pande, P.~Riley, {\it J. Comput.-Aided
  Mol. Des.\/} {\bf 30}, 595 (2016).

\bibitem{wu2018moleculenet}
Z.~Wu, {\it et~al.\/}, {\it Chem. Sci.\/} {\bf 9}, 513 (2018).

\bibitem{houchins2017quantifying}
G.~Houchins, V.~Viswanathan, {\it Phys. Rev. B\/} {\bf 96}, 134426 (2017).

\end{thebibliography}

\bibliographystyle{Science}

\section*{Acknowledgments}
We thank Matt Wolski of Daramic for providing us with polyporous separator samples. This material is based upon work supported by the National Science Foundation under Grant No. 1944007. Funding for this research was provided by the Abdul Latif Jameel World Water and Food Systems Lab (J-WAFS) at MIT. N.L. acknowledges support by the National Science Foundation Graduate Research Fellowship under Grant No. 1122374.  D.K. and V.V. gratefully acknowledge funding support from the National Science Foundation under award CBET-1554273. D.K and V.V. thank Bharath Ramsundar for useful discussions and feedback about the computational models and the deep learning methodology. V.V. acknowledges support from the Scott Institute for Energy Innovation at Carnegie Mellon University. D.K. acknowledges discussions with Victor Venturi regarding the deep learning model implementation.\\
  
\section*{Author Contributions}
Conceptualization, N.L. and K.M.; Methodology - Experimental - N.L.; Methodology - Modeling - D.K., V.V.; Investigation, N.L. and M.L.G.; Formal Analysis, D.K. and V.V.; Data Curation, D.K.; Writing - Original Draft, N.L. and D.K.; Writing - Review \& Editing - N.L., D.K., K.M., and V.V.; Supervision. K.M. and V.V. \newline

\section*{Competing Interests}
D.K., V.V., N.L., and K.M. are inventors on a provisional patent application, 63/066841, related to hydrogen donors for lithium-mediated ammonia synthesis.
\newline

\section*{Data and Materials Availability}
 All code and supporting data used in the work will be made available on Github. We will release this dataset into the MoleculeNet benchmark suite named as the KamletTaft-dataset.
 \newline 
\section*{Supplementary materials}
\vspace{-0.5cm}
\hspace{2pt}\\
Materials and Methods\\
Supplementary Text\\
Figs. S1 to S11\\
Tables S1 to S5\\
References \textit{(1-17)}

\newpage

\begin{figure}[H]
		\centering
\includegraphics[width=\linewidth]{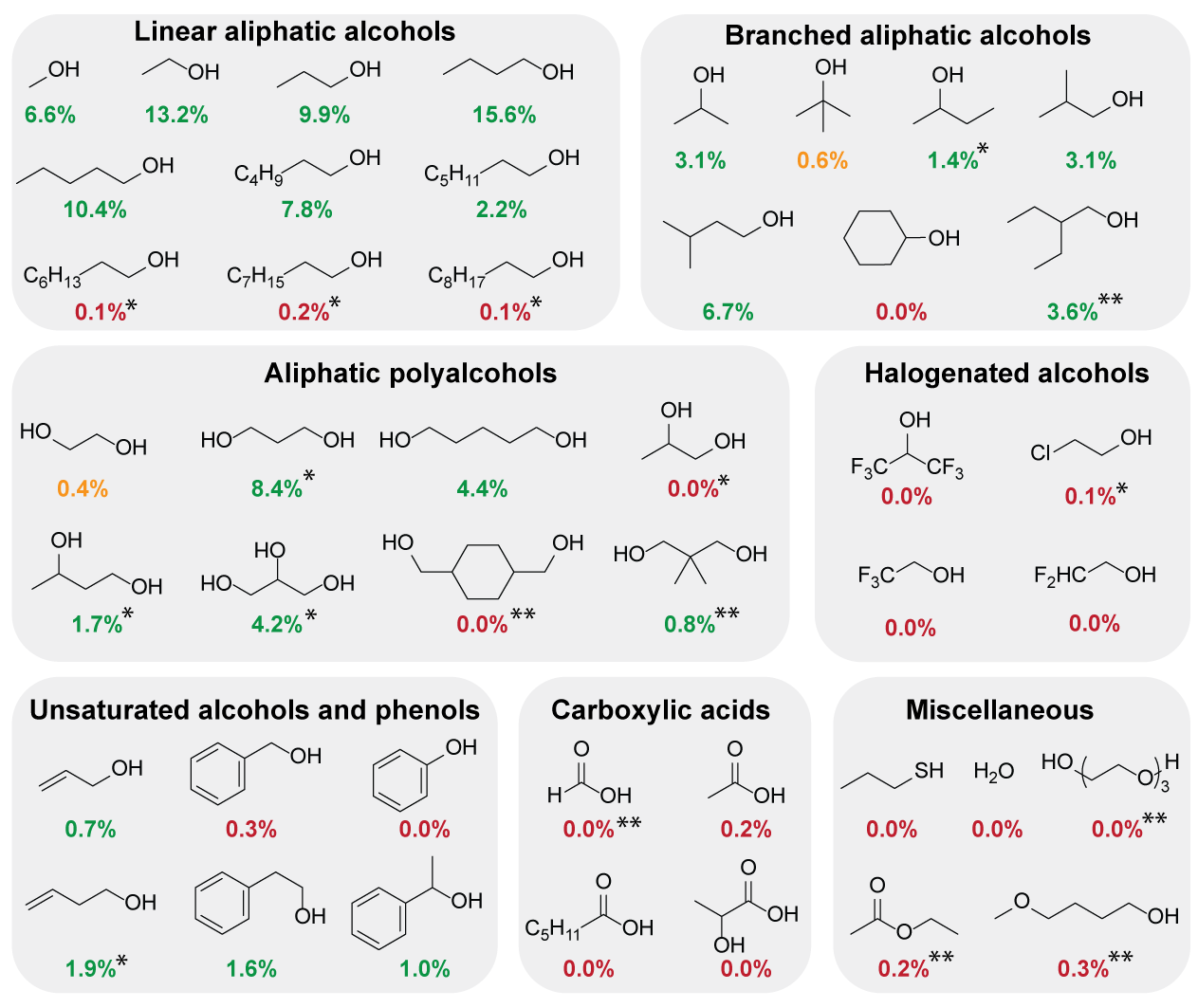}
		\caption{\textbf{Maximum obtained ammonia Faradaic efficiencies for a variety of tested proton donors.} Proton donors for which FE values are given in green are classified as active (ammonia FE > 0.5\%), those in red are classified as inactive (ammonia FE < 0.5\%). One proton donor, tert-butanol, was classified as inactive as the maximum obtained FE (labeled in yellow) did not exceed 0.5\% when accounting for the error in the measurements. Note that the conditions at which maximum reported FEs were obtained differ between proton sources (Table S2). Proton donors labeled with a star (*) were used in closed-loop improvement of an interpretable model (see below), while those labeled with two stars (**) were selected for validation of a deep-learning model (see below).} 
		\label{fig2:FEs}
\end{figure}

\begin{figure}[H]
		\includegraphics[width=1.1\linewidth]{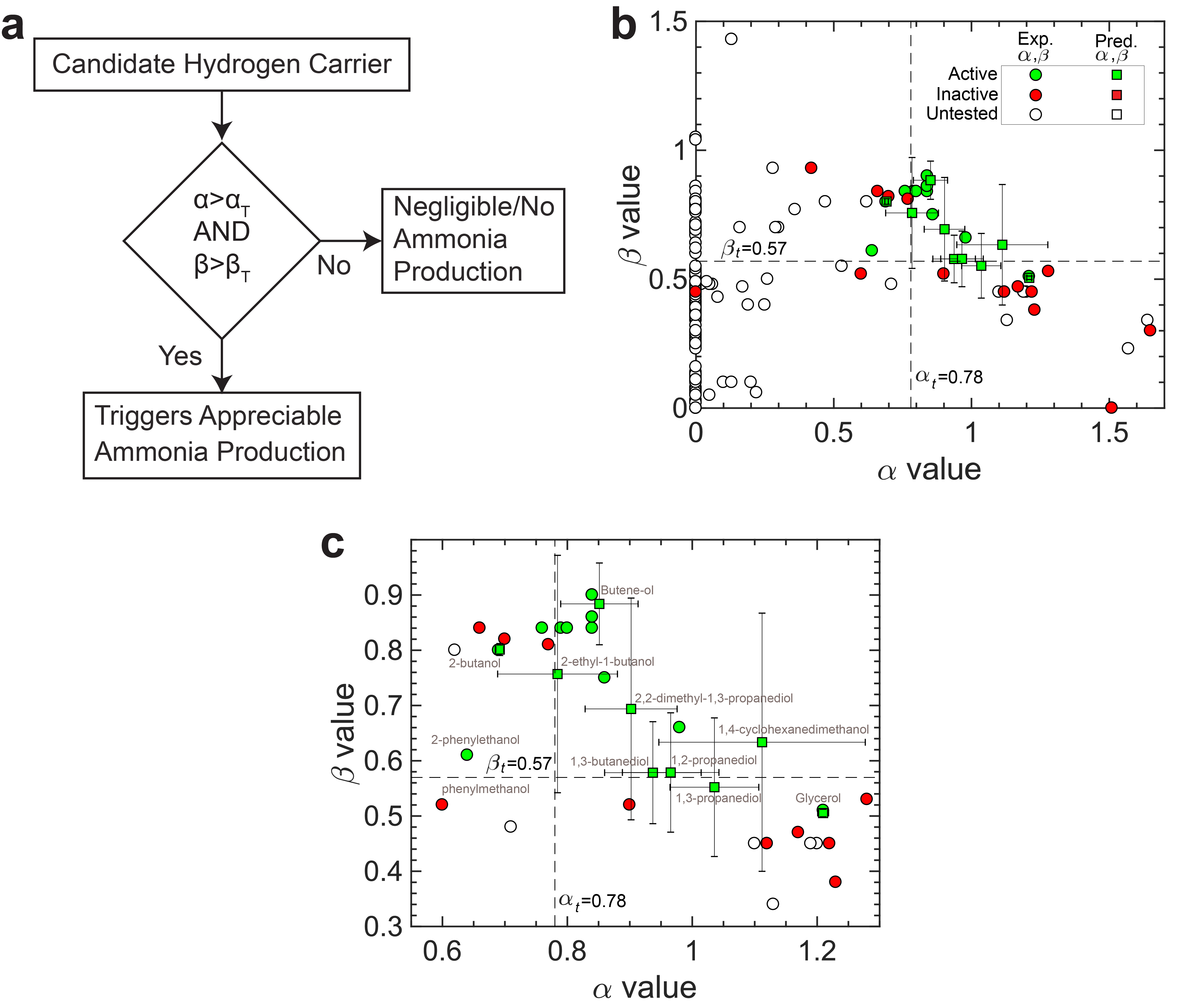}
		\caption{\textbf{Interpretable classification model to identify molecular descriptors of activity towards ammonia production.} (a) A decision tree that has a high classification accuracy ($\approx$ 95\%) and identifies Kamlet-Taft parameters, which quantify hydrogen-bond donor and acceptor abilities (denoted as $\alpha$ and $\beta$, respectively), as the most indicative of the ability to yield ammonia. Associated threshold values for the ability to trigger ammonia production are given by $\alpha_t=0.78$ and $\beta_t=0.59$. (\textbf{b}) A range of proton donors plotted in the $\alpha-\beta$ space with either experimentally measured \cite{marcus1993properties,meyer1995correlation,Sten} or predicted from the developed deep learning model parameter values. Black dashed lines correspond to $\alpha_t$ and $\beta_t$, showing the desired quadrant for promising protons donors. Error bars along the two axes represent one standard deviation from the ensemble of prediction models. (\textbf{c}) A smaller section of $\alpha$-$\beta$ space with several measured candidates annotated.}
		\label{fig3:classModelPred}
\end{figure}

\begin{figure}[H]
		\centering
\includegraphics[width=\linewidth]{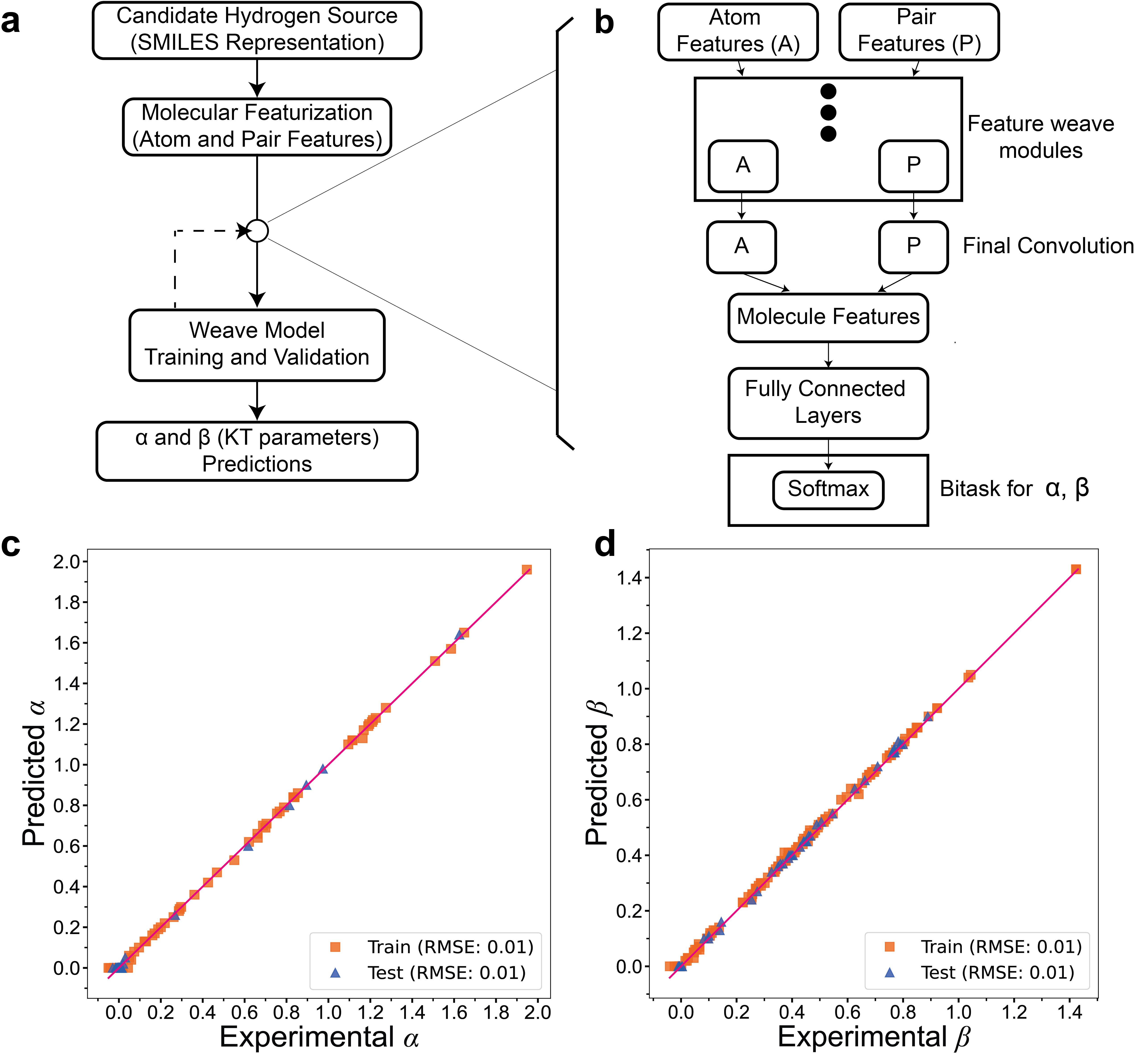}
		\caption {
		A deep learning model to predict Kamlet-Taft parameters. (\textbf{a}) Steps involved in the approach to predicting $\alpha$ and $\beta$ (KT) parameters. (\textbf{b}) The weave featurization technique and deep learning framework involving an ensemble of models for robust predictions \cite{wu2018moleculenet}. Parity plots for (\textbf{c}) $\alpha$ and (\textbf{d}) $\beta$ values obtained from the developed deep-learning model. Note that the predictions on test set after cross-validation have comparable performance to that on the training set indicating generalizability of the model.}
		\label{fig4:deepLearningPred}
\end{figure}

\begin{figure}[H]
		\centering
\includegraphics[width=\linewidth]{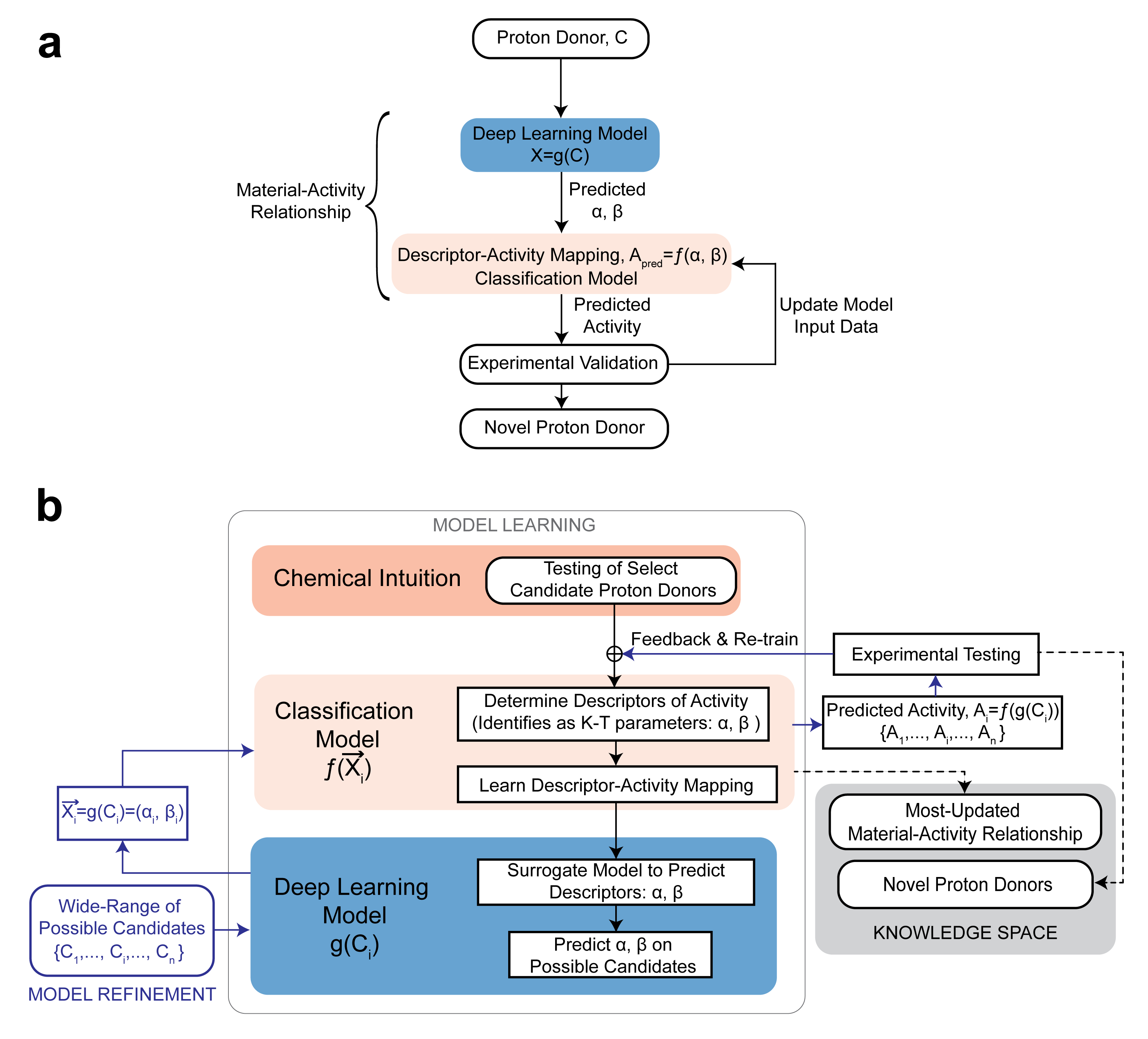}
		\caption{\textbf{Closed-Loop Learning of the Material-Activity Mapping.} (\textbf{a}) The two-part model, consisting of the interpretable decision tree model and the deep-learning model, used to proposed the next batch of experiments to test in order to learn the most about the material-activity mapping with every successive batch of experiments. (\textbf{b}) A schematic showing information flow towards identifying novel proton donors and learning the material-activity relationship. The depicted task of \textit{model learning} refers to the process of learning the material-activity relationship based on the initial set of experiments (initial model). The task of model refinement begins with activity predictions on several possible candidates using the initial model, which is then used to propose a sequence of experimental testing. The testing outcomes are used to augment the data to re-learn the descriptor-activity relationship, which closes the training loop.}
		\label{fig5:closedLoop}
\end{figure}

\begin{figure}[H]
		\centering
\includegraphics[width=\linewidth]{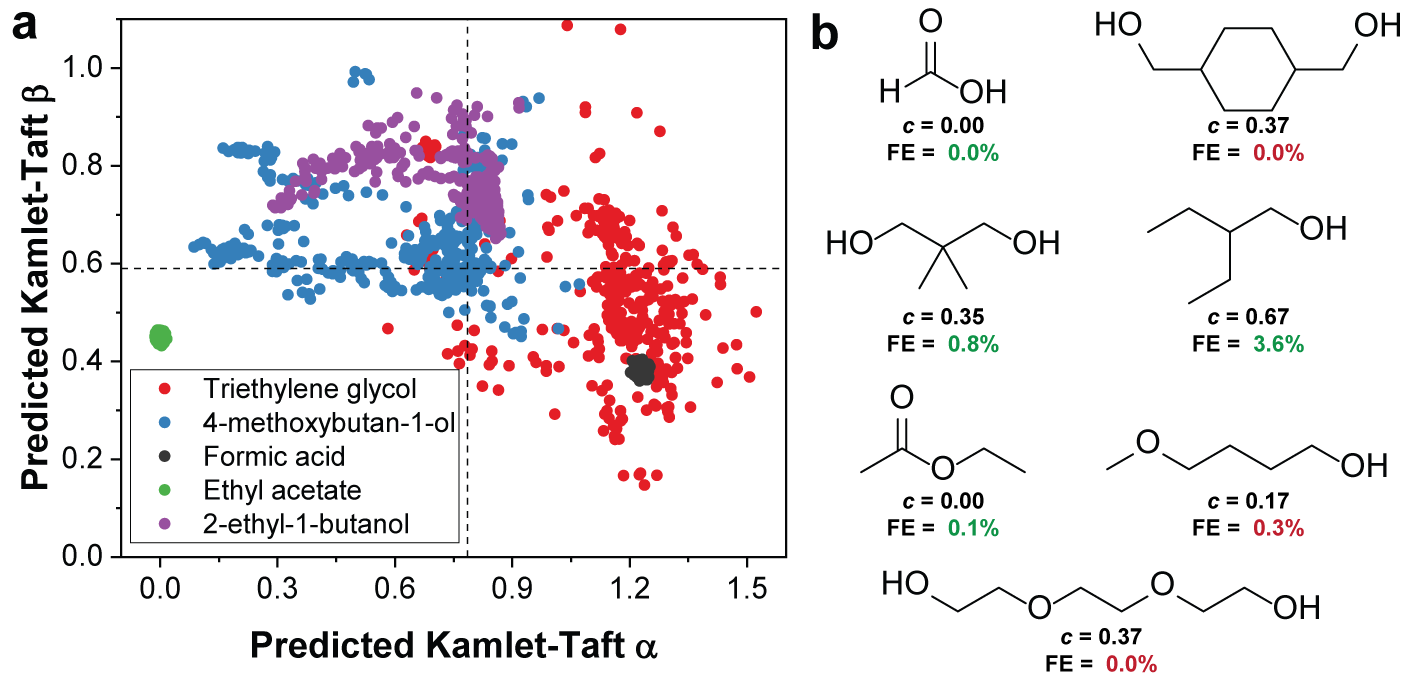}
		\caption{\textbf{Experimental testing of candidates suggested from deep-learning models.} (\textbf{a}) Predicted $\alpha$ and $\beta$ values from an ensemble of models for selected proton donors. Not all proton donors are included so as to maintain clarity in the figure. (\textbf{b}) Experimentally measured maximum FEs towards NH$_3$ for several proton donors with their $c$-values for activity. FEs given in green represent that the binary activity classification matches the predicted one, while a those in red do not match the predicted activity. 4/7 of the predictions were correct.}
		\label{fig6:expTesting}
\end{figure}

\clearpage


\section*{Supplementary material}

# Supplementary Information for Closed-Loop Design of Proton Donors for Lithium-Mediated Ammonia Synthesis with Interpretable Models and Molecular Machine Learning

Dilip Krishnamurthy,[1†] Nikifar Lazouski,[2†] Michal L. Gala, [2] Karthish Manthiram[2*] and Venkatasubramanian Viswanathan[1*]

[1]Department of Mechanical Engineering,
Carnegie Mellon University, Pittsburgh, PA, USA
[2]Department of Chemical Engineering,
Massachusetts Institute of Technology, Cambridge, MA, USA

†Equally Contributing Authors; *E-mails: karthish@mit.edu, venkvis@cmu.edu

## Materials

Tetrahydrofuran (THF, 99+%, stabilized with BHT), molecular sieves (3Å, 4-8 mesh), 1-propanol (99+%, extra pure), L-alanine (99%), and 2-methyl-1-propanol (isobutanol, ACS reagent, spectro grade, 99+%) were purchased from Acros Organics. Lithium tetrafluoroborate (LiBF$_4$, 98%), tert-butyl alcohol (99%), 2-butanol (>99%), 2-ethyl-1-butanol (98%), 1-pentanol (ACS reagent, ≥99%), 1-hexanol (reagent grade, 98%), 1-heptanol (98%), 1-nonanol (98%), benzyl alcohol (99.8%, anhydrous), phenol (unstabilized, ≥99%), 1-phenylethanol (98%), 2-phenylethanol (99%), 2-chloroethanol (99%), 2,2,2-trifluoroethanol (ReagentPlus, ≥99%), hexafluoro 2-propanol (≥99%), ethylene glycol (anhydrous, 99.8%), 1,3-butanediol (±, 99%, anhydrous), glycerol (ReagentPlus, ≥99%), triethyleneglycol (ReagentPlus, 99%), 1,5-pentanediol (≥97%), acetic acid (ReagentPlus, ≥99%), hexanoic acid (≥99%), allyl alcohol (99%), 2-methoxyethanol (99.8%, anhydrous), 1-propanethiol (99%), hydrochloric acid (HCl, ACS Reagent, 37%), sodium salicylate (ReagentPlus, ≥99.5%), and sodium hypochlorite (NaOCl, 10-15%) were purchased from Sigma-Aldrich. Methanol (anhydrous, 99.9%), cyclohexanol (99%), 3-butene-1-ol (98+%), sodium nitroprusside (99-102%), and ammonium chloride (NH$_4$Cl, anhydrous, 99.99%) were purchased from Alfa Aesar. Ethanol (Koptec, anhydrous, 200 proof), 2-propanol (Semi grade, BDH), sodium hydroxide (NaOH, Macron Fine Chemicals, pellet

1

form), and acetone (ACS, BDH Chemical) were purchased from VWR International. 1-butanol (Certified ACS), 3-methyl-1-butanol (isoamyl alcohol, for molecular biology), dichloromethane (DCM, 99.5%), and hexanes (C6H14) were purchased from Fisher Scientific. Formic acid (ACS Reagent, 98-100%) and ethanolamine were purchased from EMD Millipore. Milli-Q water was obtained by filtering deionized (DI) water through a Milli-Q purification system (Merck, Millipore Corporation). Platinum foil (Pt, 0.025 mm thick, 99.99%, trace metals basis) and 1-octanol (99%) were purchased from Beantown Chemical. Argon gas (UHP, 5.0 grade) was purchased from Airgas. Nitrogen gas was available in-house; it is generated by boil-off of liquid nitrogen from Airgas. Steel foil (cold-worked 304 stainless steel, 0.002" thick) was purchased from McMaster-Carr. Polyporous Daramic 175 separators were received as a sample from Daramic (Charlotte, NC).

## Electrolyte preparation

Dry molecular sieves were prepared by washing as-purchased or previously used molecular sieves with acetone and drying in a muffle furnace at 300 °C for 5 hours. The sieves were added as 20% by volume to as-purchased THF in a round-bottom flask. The flask was sealed from the atmosphere with a rubber septum are dried for at least 96 hours before use.

As purchased $LiBF_4$ was dissolved in dry THF to obtain a 1 M $LiBF_4$ in THF electrolyte solution. The $LiBF_4$ must be sufficiently pure for successful ammonia production; we found that salt purchased from Sigma-Aldrich is sufficiently pure for these experiments, while other vendors' may require additional purification; one potential purification procedure is given in prior work (*1*). The solution was centrifuged at 6000 rpm (4430 rcf) to remove insoluble precipitates. The clear solution was transferred to oven-dried vials, stored in a desiccator, and used within 12 hours of preparation. The solution transfer operations can be performed in the ambient atmosphere; the solutions should not be stored open to the atmosphere, however, as the electrolyte solution can absorb a significant amount of water from ambient air.

The proton donor was added to the electrolyte immediately prior to experiments. The total volume of proton donor-containing electrolyte solution prepared for each experiment is 4 mL. If the volume of proton donor that needs to be added to obtain the desired concentration is <100 $\mu$L, then the proton donor was added to 4 mL of electrolyte directly. If the volume required is >100 $\mu$L, then the proton donor added to a smaller volume of electrolyte that was rounded to the nearest 0.1 mL, so that the final volume of the proton donor in electrolyte solution would equal 4 mL. For example, to prepare 0.2 M ethanol, 47 $\mu$L of ethanol were added to 4 mL of electrolyte solution, while to prepare 0.6 M 1-butanol, 220 $\mu$L of 1-butanol were added to 3.8 mL of electrolyte solution.

## Nitrogen reduction experiments

Polished stainless steel electrodes were used as the cathode Stainless steel shims were cut into 2x2 cm pieces, wet with DI water, and polished with 400 grit followed by 1500 grit sandpaper thoroughly. The polished foils were rinsed thoroughly with DI water and dried in air at 80 °C. Stainless steel cathodes were used in a single experiment before discarding.

Parallel plate cells described in prior work were used to perform nitrogen reduction experiments (*1*). Briefly, a polished steel foil was used as the cathode, a platinum foil was used as the anode, a piece of Daramic was used as a separator, and machined polyether ether ketone (PEEK) cell parts were used for the cell body (Fig. S1). All cell parts were dried in air at 80°C for at least 20 minutes prior to use.

Nitrogen (or argon, in control experiments) gas was flowed at 10 standard cubic centimeters per minute (sccm) through a vial containing THF and molecular sieves to saturate the feed gas with THF. The THF-saturated feed gas was then flowed to an assembled 2-compartment cell. The proton-donor electrolyte was added first to the anode compartment, then to the cathode compartment. 1.75 mL of electrolyte was added to each compartment; note that this is the volume added to each compartment, and may not be the final volume in each compartment at the conclusion of the experiment due to solvent evaporation. The feed gas was flowed through the electrolyte for 10 minute at open circuit to saturate the electrolyte with gas and to strip oxygen from the solution.

After saturating the solution with the feed gas, a constant current of 20 mA was applied for 6 minutes using a Tekpower 5003 DC power source, for a total of 7.2 coulombs of charge passed. In some experiments (see Supplementary Table 2), the potential required to applied 20 mA exceeded 50 V. In experiments where an excess of 50 V was required to apply 20 mA, a constant potential of 50 V was applied across the cell for 6 minutes, and the total charge passed was quantified by measuring the potential drop across a resistor in series with the cell (Fig. S6). As the electrolyte resistance does not significantly change with concentration of most proton donors, the higher voltage required is likely due to changes in SEIs at the cathode or anode in these experiments.

Following application of current, the catholyte was immediately removed from the cell and diluted in water. In most experiments, the electrolyte was used directly to prepare samples for ammonia quantification. In these cases, the samples were made as follows: one by adding 200 $\mu$L of catholyte to 1800 $\mu$L of Milli-Q water (10-fold dilution), and another by adding 100 $\mu$L of catholyte to 1900 $\mu$L of water (20-fold dilution), to be able to accurately quantify ammonia at both lower and higher Faradaic efficiencies. In some cases, the proton donor can affect the colorimetric assay negatively by either phase separating with water (e.g. octanol), leading to higher spurious absorbances, or chemically (e.g. ethyl acetate, thiols), leading to

lower or shifted absorbances (Fig. S3). In these cases, the proton donor was extracted from the ammonia-containing samples. To extract the proton donor, 500 $\mu$L of electrolyte were added to 4.5 mL of 0.05 M $H_2SO_4$ in water. The proton donor in resulting acidified solution was extracted with 3 mL of either DCM or hexanes three times. Milli-Q water was then added to the aqueous phase to a final volume of 5 mL if the volume decreased, which may occur if the THF was extracted into the organic phase. The aqueous phase was centrifuged at 6000 rpm (4430 rcf) for 10 minutes to promote complete phase separation. The aqueous phase was then quenched with base by adding 1500 $\mu$L of the acidified solution to 500 $\mu$L of 0.4 M NaOH, or by adding 750 $\mu$L of the acidified solution to 250 $\mu$L of 0.4 M NaOH and 1000 $\mu$L of Milli-Q water.

After experiments, Daramic separator pieces were rinsed with acetone and soaked in DI water for at least 10 minutes to remove traces of solvent and ammonia. Cell parts and platinum anodes were rinsed with acetone and washed thoroughly with DI water. All cell parts, electrodes, and separators were dried at 80 °C in air prior to use in further experiments.

## Choosing proton carrier concentrations

In order to determine whether a proton donor can be used to produce ammonia using the lithium-mediated approach, a range of proton concentrations had to be efficiently screened for activity. From prior work (*1*), it is known that ammonia yields depend on the concentration of ethanol, the proton donor. At low concentrations, no ammonia is formed and a large amount of lithium remains on the cathode, while at high concentrations, no ammonia is formed due to competition from the hydrogen evolution reaction. We posited that this behavior is not unique to ethanol and can be observed for various proton donors. From this hypothesis, we developed a heuristic to rapidly screen the concentration range. Initially, electrolyte containing 0.2 M of the proton donor was used to test for ammonia production. If a significant amount of lithium metal was found to remain on the cathode or in solution, the concentration of the proton donor was increased for the next run. Typically, the concentration was increased 2- or 3-fold, depending on the extent of lithium coverage on the surface. If the surface is clean and the steel cathode is visible, the concentration of the proton donor was decreased by a similar amount. Every proton donor was tested at three concentrations, with a maximum concentration tested of 1 M. The 1 M cutoff is arbitrary, and was chosen as it is a concentration that results in fairly large volume fractions of proton donor in electrolyte for most proton donors. Using this heuristic, we typically obtained runs with a high lithium coverage after the experiment, a low lithium coverage, and an intermediate coverage, except for cases where 1 M of proton donor did not decrease lithium coverage; in these cases, all experiments had a large lithium coverage. We believe that approach allowed us to probe a large compound-concentration phase space efficiently to determine which compounds are capable of promoting lithium-mediated nitrogen reduction.

## Ammonia quantification

The amount of ammonia in samples produced in nitrogen reduction experiments was quantified by using the salicylate assay according to a procedure described in earlier work (*1*). Briefly, 280 $\mu$L of 1% NaOCl in 0.4 M NaOH solution was added to 2 mL of ammonia sample solution, followed by 280 $\mu$L of 2.5 M sodium salicylate, 3.5 mM sodium nitroprusside solution. The resulting solution was mixed vigorously and left to evolve color in the dark for at least 2 hours. The absorbance spectrum of the resulting solution was measured using an Ocean Optics Flame-S UV-Vis spectrophotometer. The relevant signal for ammonia quantification was taken to be the difference in absorbance values at 650 nm and 475 nm to avoid overestimating the amount of ammonia produced (*1*).

A fresh ammonia calibration curve was made for each quantification batch. Calibration curves were made by adding 100 $\mu$L of electrolyte to solutions of known ammonium sulfate concentrations, ranging from 0 to 80 $\mu$M. A typical calibration curve and absorbance spectra can be seen in Fig S9. Addition of most proton donors to the electrolyte used do not change the calibration curve significantly; proton donors which may affect the quantification (such as thiols or long chain alcohols) were typically removed by extraction prior to quantification (Supplementary Table 2). The lowest accurately quantifiable concentration of ammonia in the solutions was typically 2 $\mu$M, computed from the error in the intercept of the calibration curve. Assuming a ten-fold dilution of the electrolyte solution to make a sample, the minimum quantifiable ammonia FE from a typical experiment is 0.1%.

## Analysis of water content

A possible explanation for the differences in activity between various compound could the difference in water content between various proton donors. However, even the most water-rich proton donor tested, triethylene glycol, contained only 2600 parts per million (ppm) water, which corresponds to a water concentration of 130 mM in the pure proton donor. Considering that the maximum concentration of proton donor used was 1 M (Supplementary Table 2), the amount of water added by the proton donor is at most 17 mM. While this is a non-negligible amount, it is similar to the amount of water present in the electrolyte initially, as measured by Karl Fischer titration. In addition, an increase in the water content by 17 mM is predicted to decrease the ammonia yields of an active proton donor, such as ethanol, by 20-30% (*1*), not eliminate it entirely. As mentioned in the main text, there is not a strong correlation between the water content of the pure proton donor and its activity toward LM-NRR (Fig. S8). All this supports the notion that the differences in activity observed between proton donors are not simply due to a difference in water content, but are instead related to the chemical structure and properties of the proton donors.

# Classification Model to Identify Promising Regimes of Candidates

Classification trees are a class of machine-learning methods for constructing models to partition data into different classes. Classification models are constructed by recursively partitioning the data space and fitting a simple prediction model within each partition. The partitioning can be represented graphically as a decision tree as shown in the main manuscript (Fig. 3). The prediction error of classification trees is typically measured in terms of misclassification cost. Classification trees are designed for dependent variables that take a finite number of unordered values. A key advantage of the tree structure is its applicability to any number of variables. In this work, we employ a binary classification tree for experimental activity classification since it is particularly well suited for this application.

## The specific classification problem

In the activity classification problem at hand, we have a training sample of many observations on a class variable $Y$ (for activity towards ammonia production) that takes values 0 or 1 (inactive or active respectively), and $p$ predictor variables, $X_1,..., X_p$. Our goal is to find a model for predicting the values of $Y$ from new $X$ values. In theory, the solution is simply a partition of the $X$ space into disjoint sets, $A_1, A_2,..., A_k$, such that the predicted value of $Y$ is $j$ if $X$ belongs to $A_j$, for $j = 0, 1$. In our case, $p = 8$ with candidate descriptors of activity towards ammonia production being acid dissociation constant (pKa), donor number (DN), Dielectric Constant ($\epsilon_r$), Kamlet-Taft parameters ($\alpha$, $\beta$, $\pi$), highest occupied molecular orbital level (HOMO), lowest unoccupied molecular orbital level (LUMO), band gap (BG), Bader volume (BV).

## Classification algorithm employed

Classification tree methods yield rectangular sets $A_j$ of the predictor variable by recursively partitioning the data set one $X$ variable at a time. Several classification tree algorithms, abbreviated as C4.5 (*2*), CART (*3*), CHAID (*4*), CRUISE (*5, 6*), GUIDE (*7*) and QUEST (*8*) have been proposed since the first published classification tree algorithm, THAID (*9, 10*).

The typical algorithm (pseudocode) for the construction of classification trees is (*11*):

1. Start at the root node

2. For each ordered variable $X$, convert it to an unordered variable $X'$ by grouping its values in the node into a small number of intervals. If $X$ is unordered, set $X' = X$.

3. Perform a chi-squared test of independence of each $X'$ variable versus $Y$ (activity classification, in this case) on the data in the node and compute its significance probability.

4. Choose the variable $X^*$ associated with the $X'$ that has the smallest significance probability.

5. Find the split set $\{X^* \in S^*\}$ that minimizes the sum of Gini indexes and use it to split the node into two child nodes. The Gini index is a generalization of the binomial variance, which is used as an impurity index. Other algorithm use entropy as the impurity index.

6. If a stopping criterion is reached, exit. Otherwise, apply steps 2–5 to each child node.

7. Prune the tree with the CART method (*3*).

**Choosing the Right Size Tree**

Decision trees have to be optimized before being used for classification of new data because the highest accuracy model could be highly complex and consist of hundreds of levels. Therefore, tree optimization implies choosing the right size of tree, which involves cutting off insignificant nodes and even subtrees. Cross-validation is a typical pruning algorithm used in practice (*12*).

**Cross-validation**

The process of cross-validation is based on optimal proportion to strike the general trade-off between tree complexity and misclassification error. With an increase in size of the tree, misclassification error typically decreases and in the case of maximum tree size, misclassification error equals 0. On the other hand, complex decision trees poorly perform on generalizability towards independent data, which is termed *true predictive power* of the tree. Therefore, the primary task is to find the optimal proportion between the tree complexity and misclassification error, which is achieved through a cost-complexity function:

$$R_\alpha(T) = R_\alpha(T) + \alpha(\tilde{T}) \longrightarrow \min_T$$

where $R(T)$ is the misclassification error of the tree $T$, $\alpha(\tilde{T})$ is complexity measure which depends on $\tilde{T}$, the total sum of terminal nodes in the tree. The $\alpha$ parameter is found through the sequence of in-sample testing when a part of learning sample is used to build the tree, the other part of the data is taken as a testing sample.

The process repeated several times for randomly selected learning and testing samples. Although cross-validation does not require adjustment of any parameters, this process is time consuming since the sequence of trees is constructed. Because the testing and learning sample are chosen randomly, the final tree may differ from time to time. The classification tree (model) reported in this work is well-converged and validated through several runs with different starting points.

## Deep Learning Model to Predict Identified Molecular Descriptors

The deep learning model's input is the molecular features of candidates obtained from the simplified molecular-input line-entry system (SMILES). Out of several molecular featurization approaches on SMILES representations, we find that the Weave featurization coupled with a Weave model (deep neural network) (*13*) yielded the most accurate (RMSE of ≈0.016 for both $\alpha$ and $\beta$) and generalizable predictions with low cross-validation error). The Weave featurization (*14*) method encodes both local chemical environment and connectivity of atoms in a molecule. The Weave featurization is similar to graph convolution in the atomic feature vectors, whereas in terms of encoding the connectivity it uses more detailed pair-wise features instead of just neighbor listing by means of "weaving" atom and pair features (Fig. 4). The Weave featurization computes a feature vector for each pair of atoms in the molecule, including bond properties, graph distance and ring information, giving rise to a feature matrix. This approach supports graph-based models that make use of properties of both atoms and bonds.

The molecular features from the weave featurization method were input into a neural network with a converged architecture, which was found to consist of two weave layers and a fully connected layer in regression mode to predict $\alpha$ and $\beta$. For the training process, the learning rate was selected to be 0.001, batch size was set to 50 and number of epochs were 100 for the 222 data points available of experimentally obtained $\alpha$ and $\beta$ values. The details of the DeepChem package, Weave featurizer and the Weave convolution model are as follows:
DeepChem version: '2.3.0'
Tensorflow version: '1.14.0'
Python version: Python 3.7.7
Featurizer: WeaveFeaturizer as implemented in DeepChem (RDKit) with default settings
Dataset Split details: frac_train=0.8, frac_valid=0.1, frac_test=0.1, with random seed
Weave Model Details/Settings (several set to default): n_tasks (number of tasks): 2, mode='regression'
n_atom_feat (number of atom features): 75
n_pair_feat (number of pair features): 14
n_hidden: 5
n_graph_feat: 128
n_weave: int = 2
fully_connected_layer_sizes: [2000, 100]
weight_init_stddevs: [0.01, 0.04]
bias_init_consts: [0.5, 3.0]
weight_decay_penalty: 0.0
weight_decay_penalty_type: "l2"
dropouts: 0.25
activation_fns: Tensorflow relu
batch_normalize: bool = True
batch_normalize_kwargs: Dict = "renorm": True,"fused": False

gaussian_expand: True
compress_post_gaussian_expansion: False.

In order to quantify the robustness of predictions we employ an ensemble of models and look for the degree of agreement between the models. The ensemble is comprised of about 400 equivalent models (404 precisely) with randomly initialized weights.

# Supplementary Figures

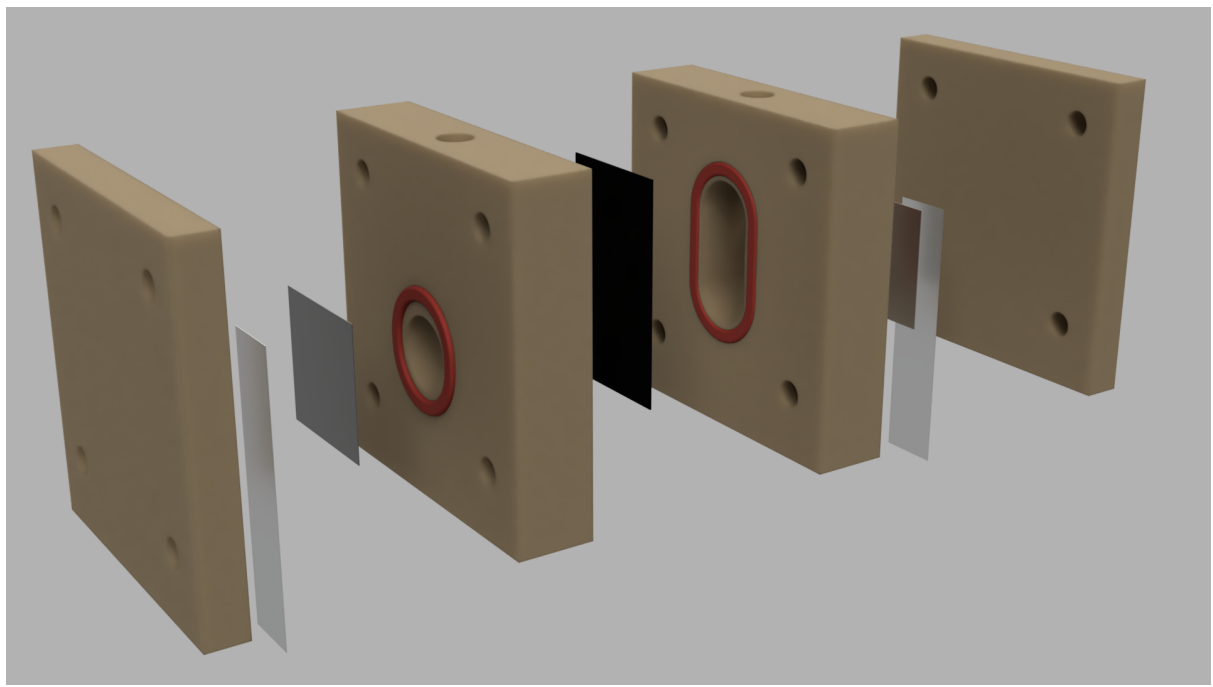

**Fig. S1. A depiction of the 2-compartment cell used in electrochemical experiments.** The cell body is made of polyether ether ketone (PEEK) polymer. Platinum and polished stainless steel foils were used as the anode and cathode, respectively. The anode and cathode compartments were separated by a piece of polyporous Daramic 175 separator. IDEX fittings were used to feed gases and plug unused holes.

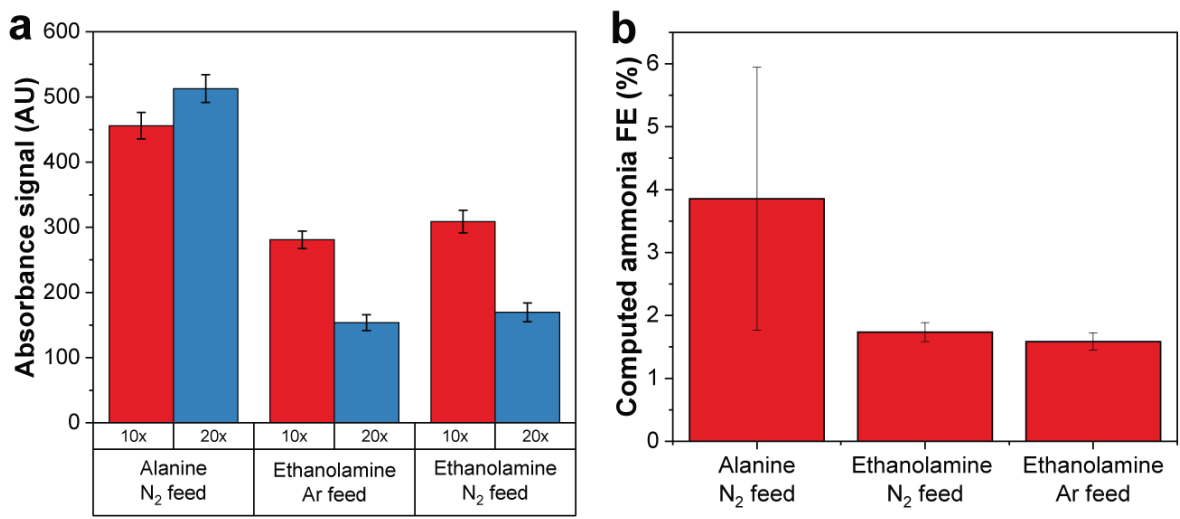

**Fig. S2. Nitrogen reduction experiments run with nitrogen-containing proton donors.** (**a**) Raw absorbance signal data for runs using 0.2 M ethanolamine or saturated alanine as a proton donor in LM-NRR. Note that alanine is not readily soluble in the electrolyte, so a saturated solution with a concentration <0.1 M was used. In these experiments, significant absorbance signals were detected in the ethanolamine argon blank solutions and alanine-containing solutions. (**b**) $NH_3$ FE values computed from the absorbance signals in (**a**). Note that a non-zero FE is computed even in the ethanolamine argon blank. The FE value computed for the alanine experiment has significant uncertainty.

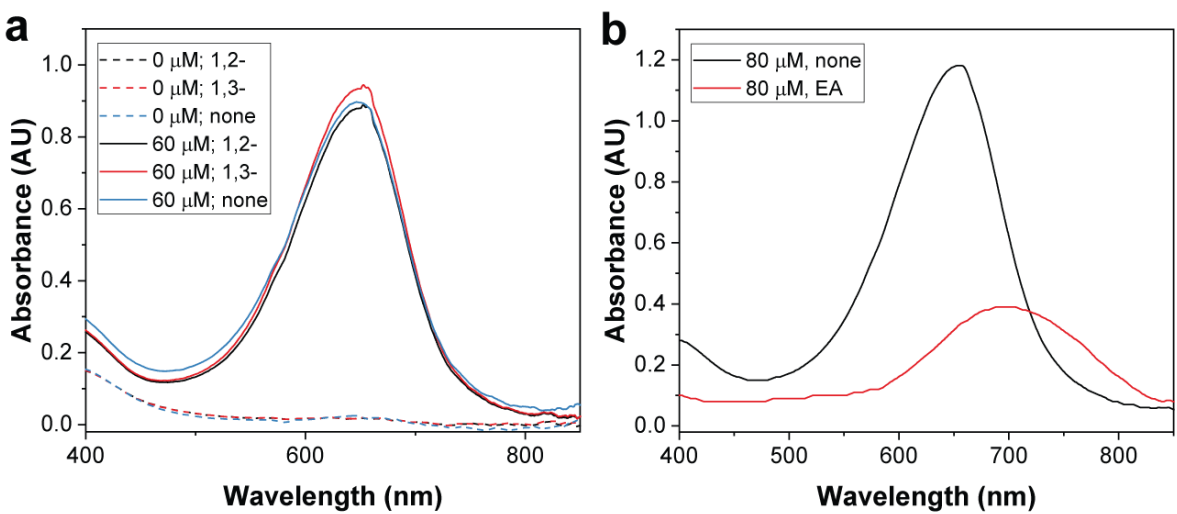

**Fig. S3. Various effects of proton donors on the salicylate quantification assay.** (**a**) Effect of addition of 0.5 M of 1,2-propanediol and 1,3-propanediol to the electrolyte used to make a 10 v/v% electrolyte in water solution containing a known concentration of ammonium. Note that the addition of the proton donors does not significantly alter the peak shape or signal magnitude. (**b**) Effect of addition of 0.5 M of 1,2-propanediol and 1,3-propanediol to the electrolyte used to make a 10 v/v% electrolyte in water solution containing 80 $\mu$M ammonium. Note that the shape of the peak changes significantly, rendering it useless for quantification of ammonia in the solution. Samples containing protons sources such as these were typically extracted with dichloromethane or hexanes prior to quantification.

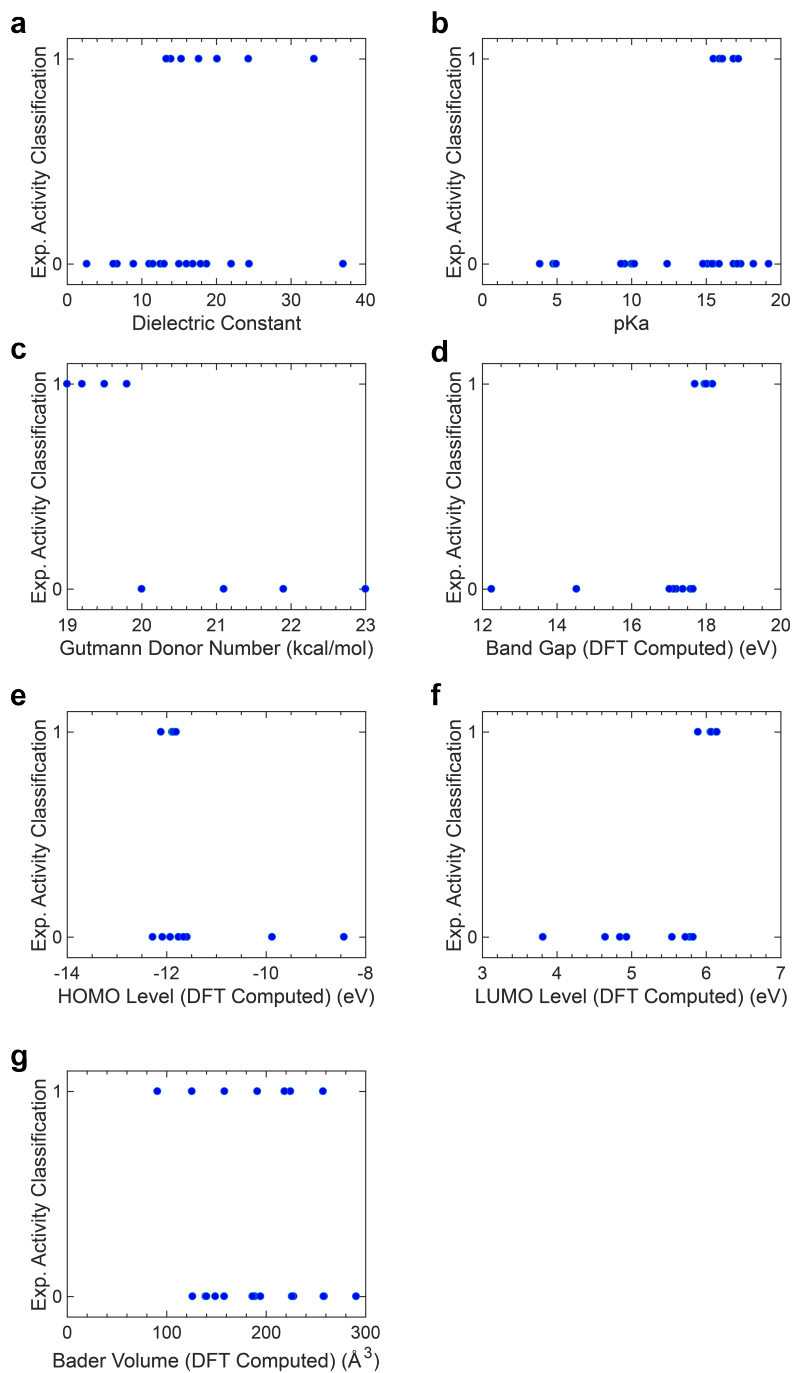

**Fig. S4. Simple refuted hypotheses for explaining experimentally proton donor activity trends.** (**a**) The binary activity of the proton donor plotted against the proton donor's pKa in DMSO. Note that while there is no direct correlation between the proton donor pKa and activity. (**b**) The binary activity of the hydrogen source plotted against the proton donor's donor number.

13

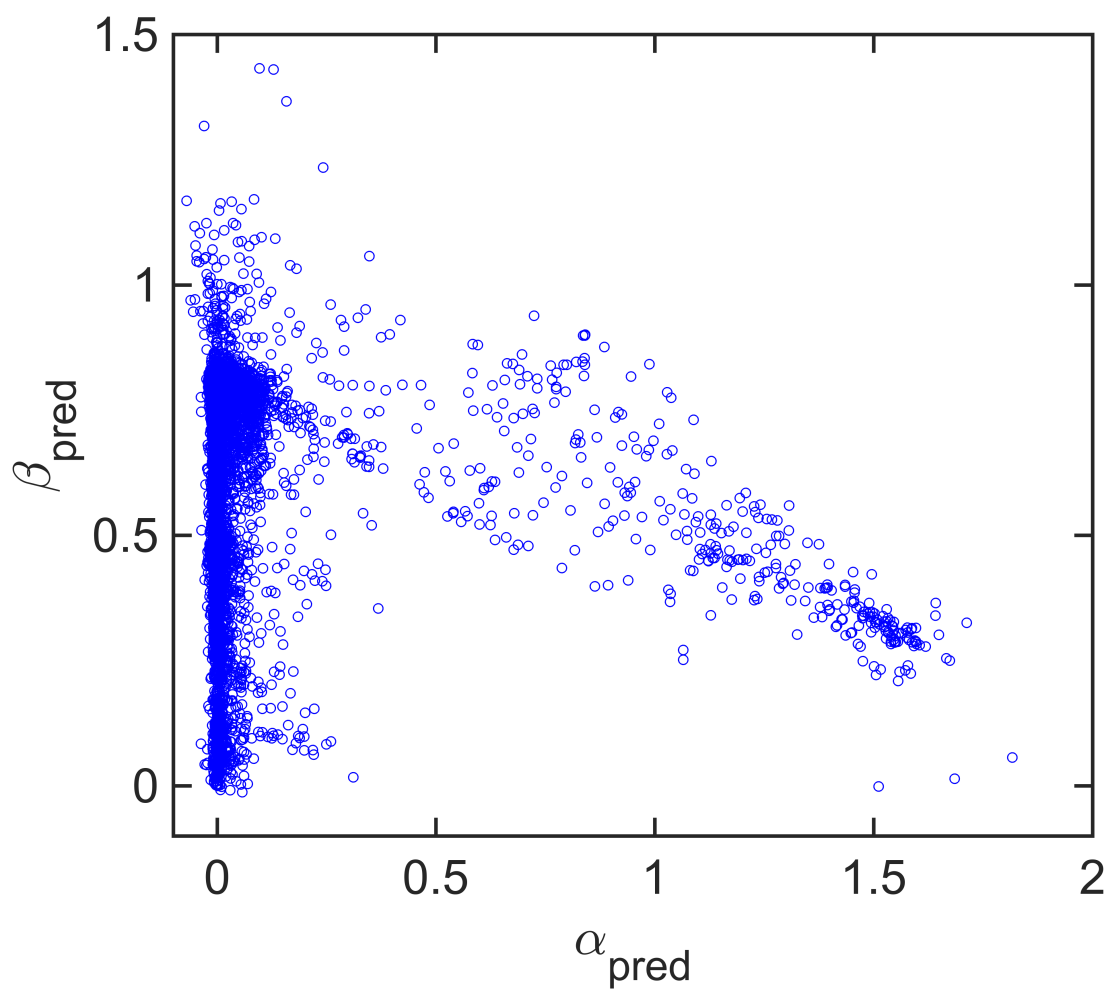

**Fig. S5. Predicted average Kamlet-Taft parameters for a subset of proton donors**. Proton donors with Pubchem IDs below 10,000 and with above-threshold (standard deviation ≤ 0.2) agreement between models in the ensemble are plotted; candidates with higher standard deviations are considered to have uncertain predictions. The inherent trade-off emerges in the $\alpha - \beta$ space as shown whereby achieving high $\alpha$ and $\beta$ is challenging, which can rationalized based on the trade-off between proton donating and accepting tendencies of molecules represented by the K-T parameters.

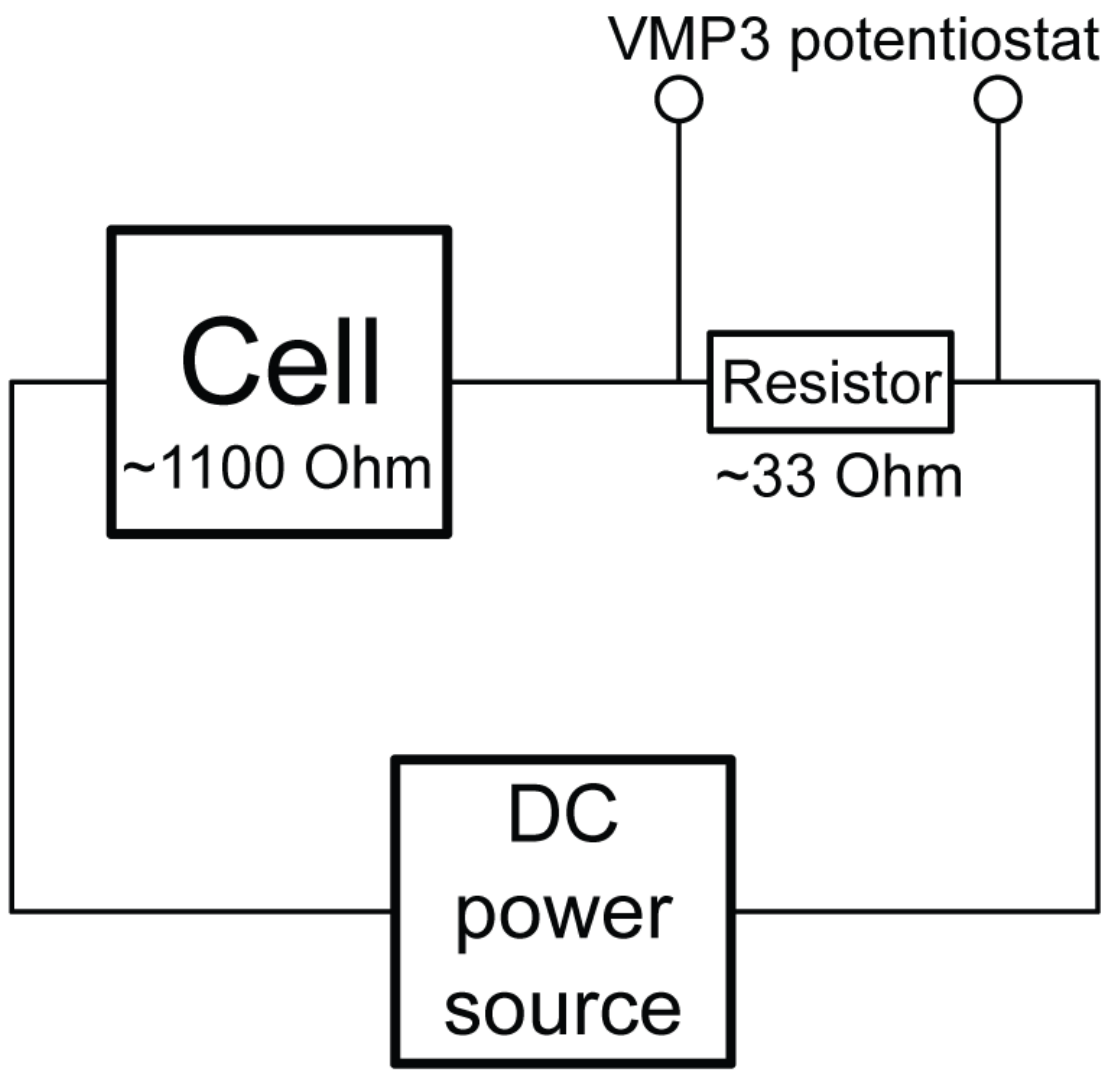

**Fig. S6. A diagram of the wiring scheme used to measure charge passed in experiments**. As the DC power source cannot independently quantify charge, the current passed through the circuit was measured with an accurate VMP3 potentiostat.

15

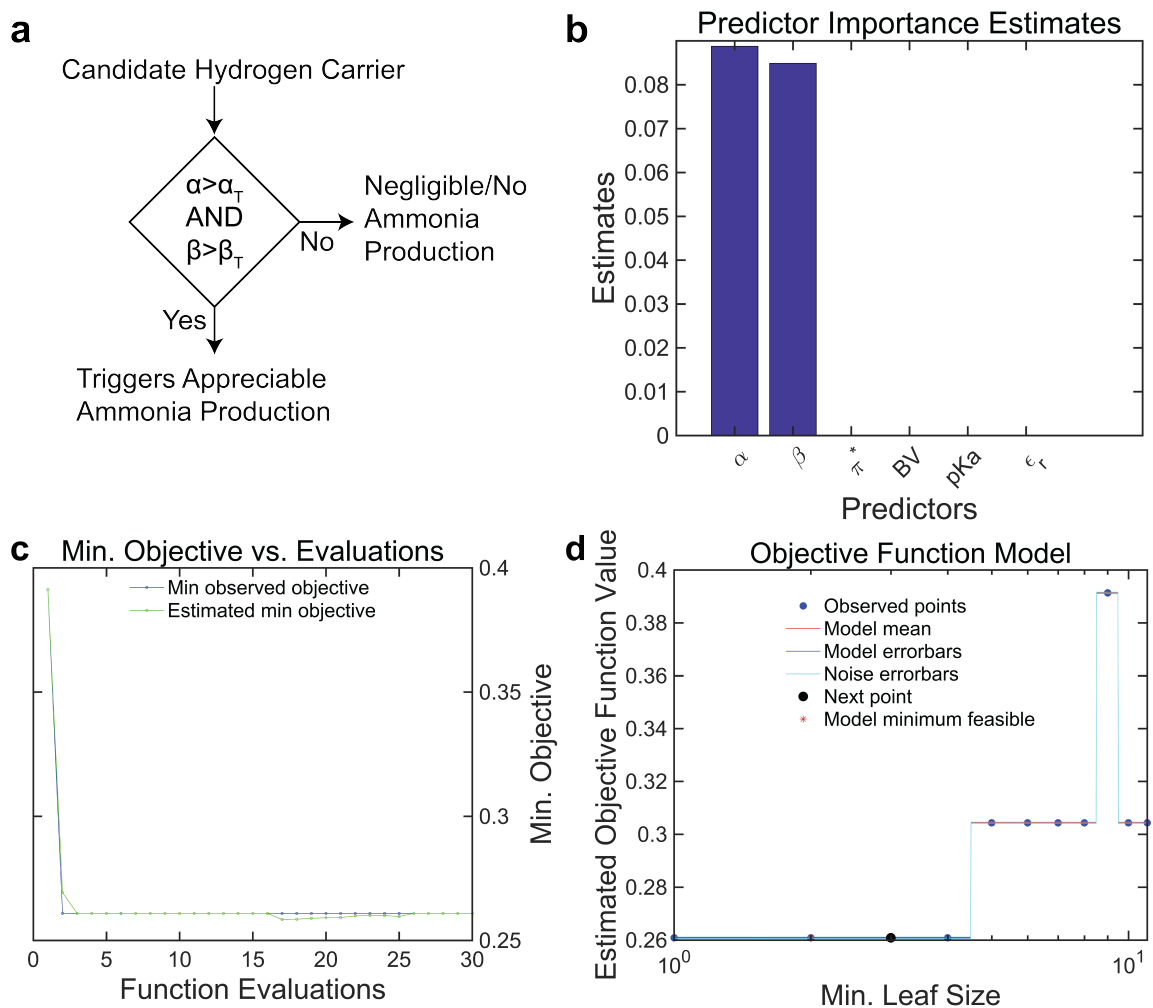

**Fig. S7. Details of the activity classification model based on decision trees.** (a) Identified classification model with $\alpha_t$ and $\beta_t$ equal to 0.815 and 0.59 respectively. (b) Predictor importance estimates by summing up changes in the risk due to splits on every predictor and dividing the sum by the number of branch node. (c) Minimum of the objective function, normalized misclassification rate, as a function of functional evaluations. (d) Evolution of the estimated objective function value with the associated minimum leaf size (classification model complexity).

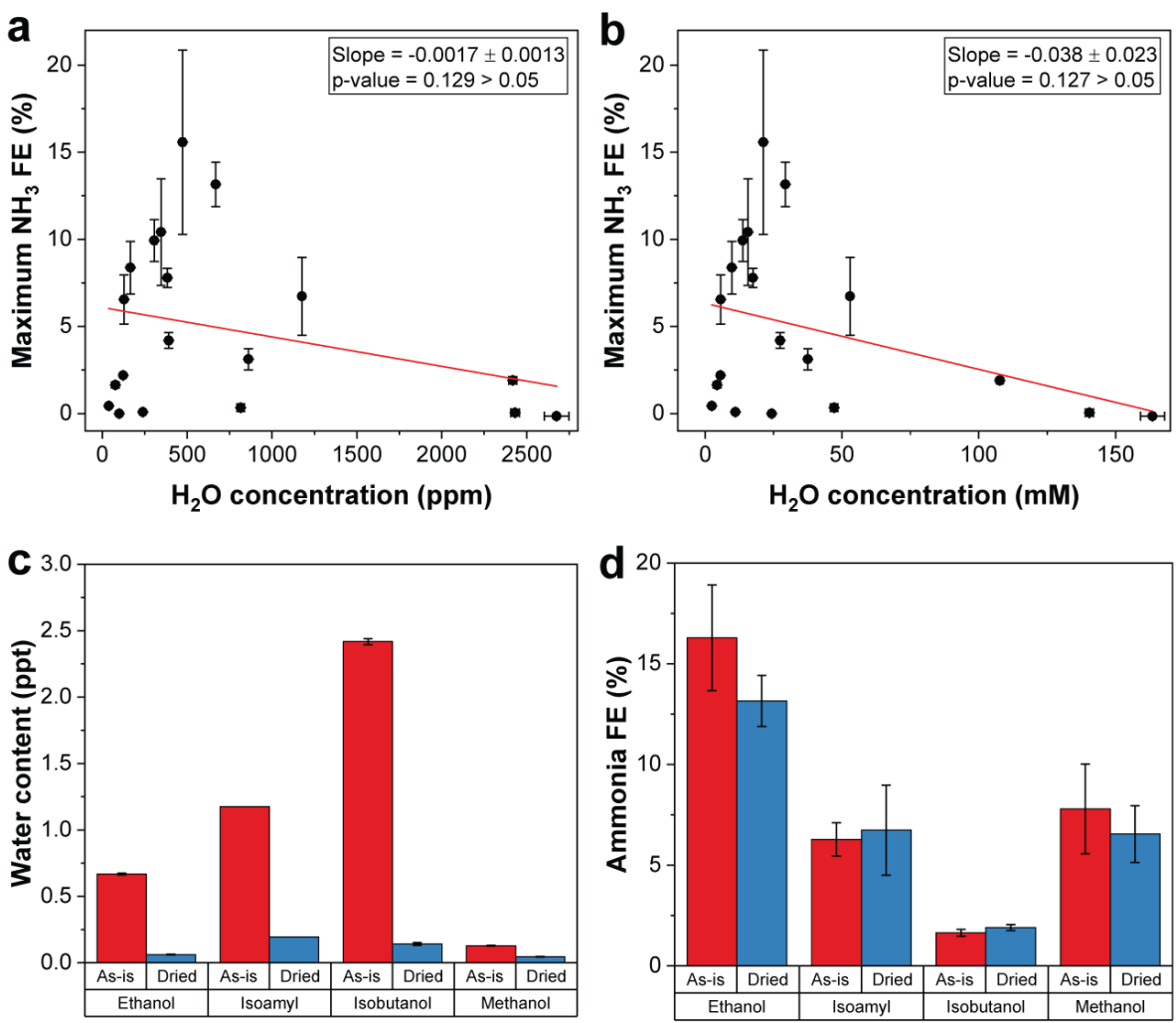

**Fig. S8. Effect of water in the proton donor on nitrogen reduction activity (a, b)** Dependence of the maximum ammonia FE obtained with a single proton donor on water content **(a)** parts per million (ppm) and **(b)** millimolar units. Note the maximum FE correlates poorly with water concentration (p≈0.12 for a non-zero slope), suggesting that the water content of the proton donor is a poor predictor of nitrogen reduction activity. **(c)** The water content in parts per thousand (ppt) of various proton donors before drying with molecular sieves and after. **(d)** The $NH_3$ FE for several proton donors before and after drying the proton donors with molecular sieves. Note that the FE does not change in a predictable manner after drying with sieves.

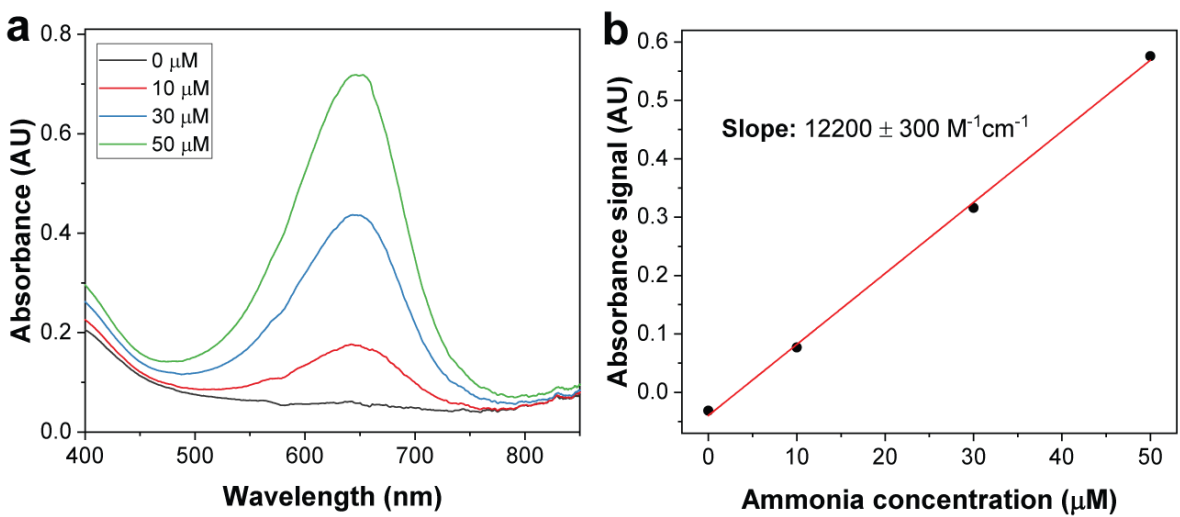

**Fig. S9. Typical ammonia quantification calibration curves.** (**a**) Absorbance spectra obtained from solutions containing known concentrations of ammonia ions in 10 v/v% 1 M LiBF$_4$ in THF electrolyte in water. (**b**) A typical calibration curve made from the spectra in (**a**). Note that the absorbance signal is taken to be the difference between the absorbance at 650 and 475 nm (*1*).

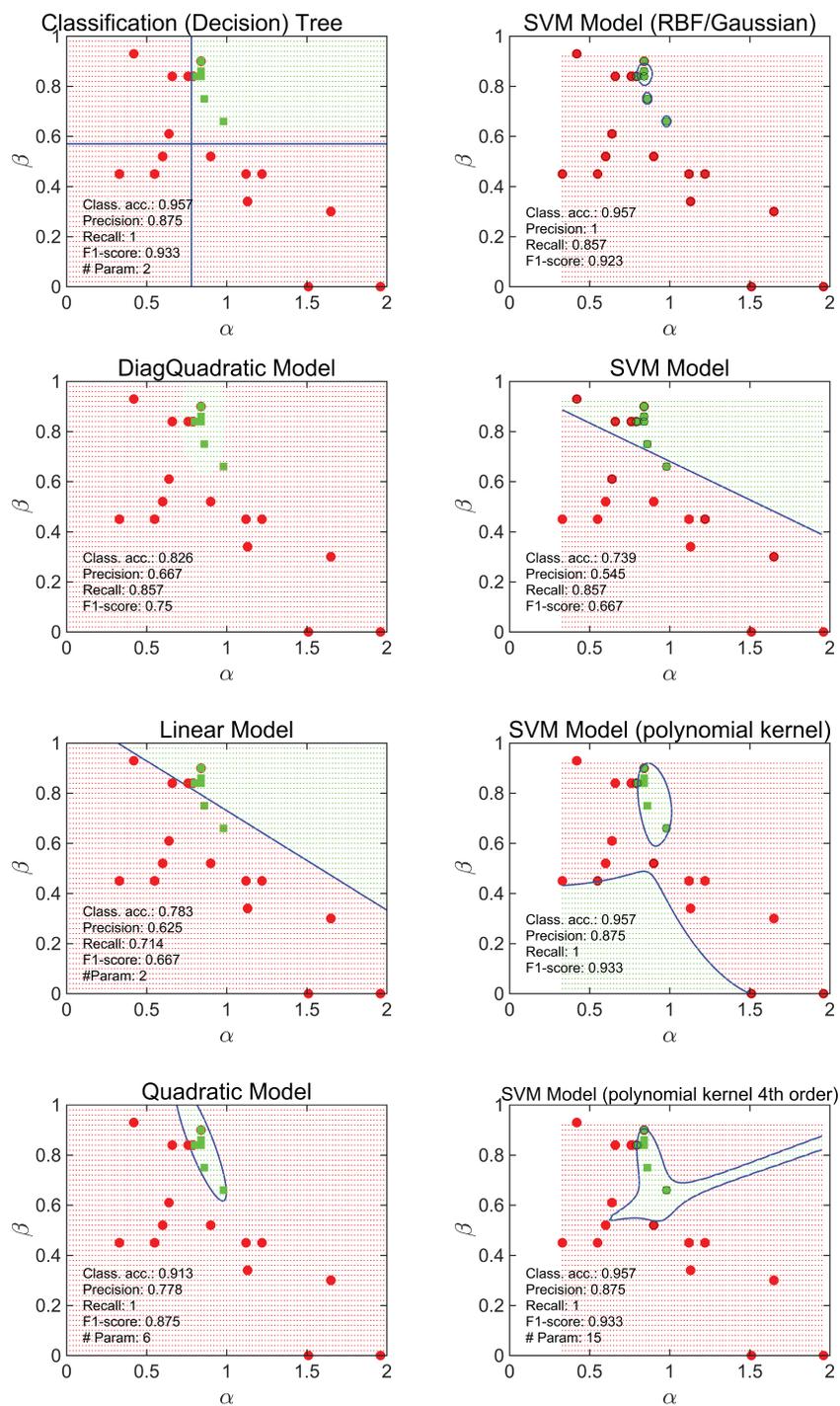

**Fig. S10.** Classification models trained on the initial set of data (post initial experimental testing) to delineate active candidates from inactive candidates.

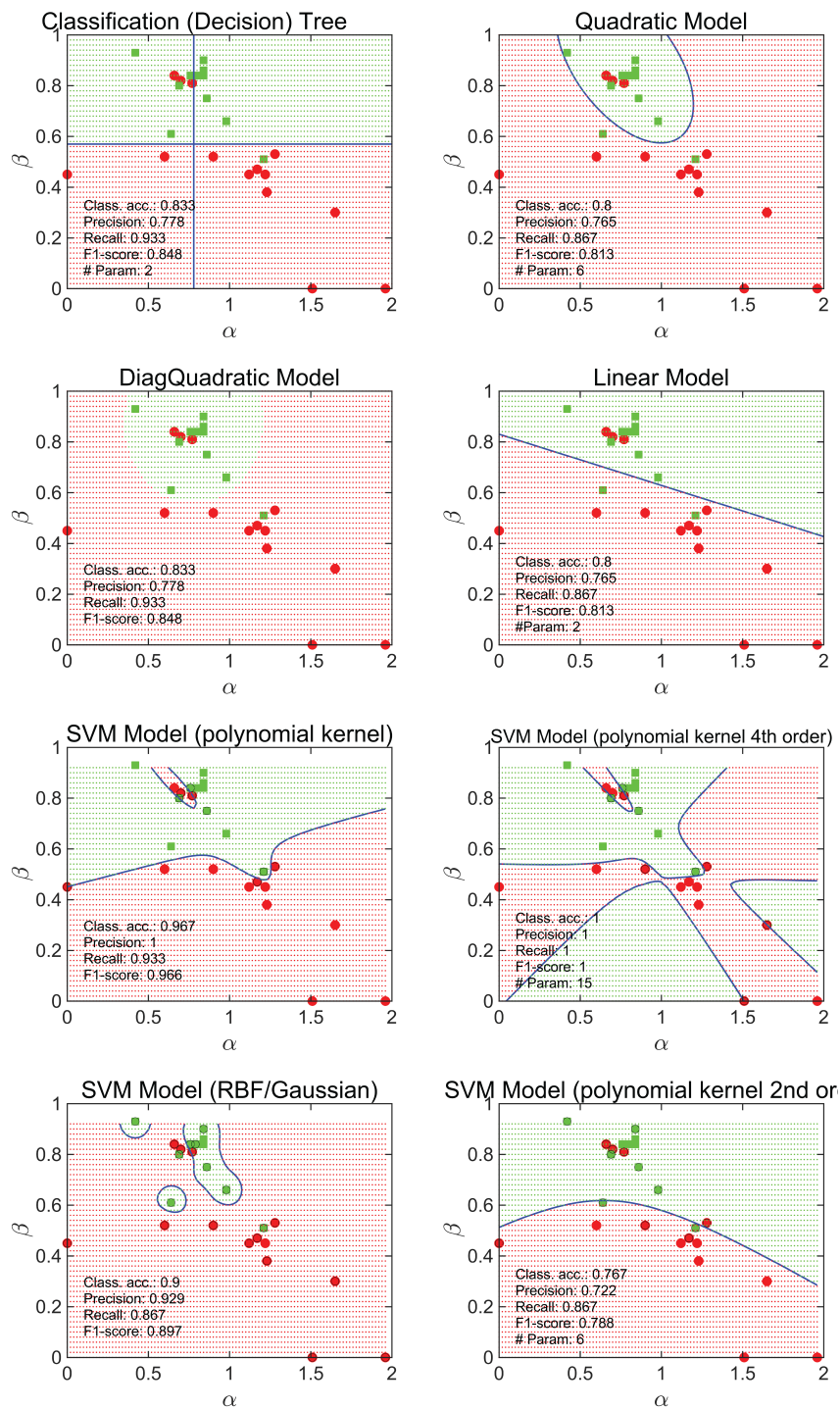

**Fig. S11.** Classification models trained on the initial set of data (after all experimental testing) to delineate active candidates from inactive candidates.

# Supplementary Tables

**Table S1. Measured water content of pure proton donors.**

| Proton donor | Water content (ppm) | Water concentration (mM) |
|---|---|---|
| 1,2-propanediol | 2430 ± 30 | 140 ± 1 |
| 1,3-butanediol | 78 ± 9 | 4.4 ± 0.5 |
| 1,3-propanediol | 166 ± 2 | 9.7 ± 0.1 |
| 1-butanol | 470 ± 10 | 21.3 ± 0.6 |
| 1-heptanol | 123.6 ± 0.9 | 5.62 ± 0.04 |
| 1-hexanol | 383 ± 5 | 17.5 ± 0.2 |
| 1-octanol | 239 ± 1 | 11.04 ± 0.05 |
| 1-pentanol | 346 ± 8 | 15.7 ± 0.4 |
| 1-propanol | 310 ± 10 | 13.7 ± 0.6 |
| 2-propanol | 860 ± 5 | 37.6 ± 0.2 |
| Benzyl alcohol | 820 ± 10 | 47.1 ± 0.6 |
| Cyclohexanol | 455 ± 8 | 24.3 ± 0.4 |
| Ethanol | 668 ± 6 | 29.3 ± 0.3 |
| Ethylene glycol | 39 ± 1 | 2.42 ± 0.07 |
| Glycerol | 390 ± 10 | 27.4 ± 0.8 |
| Isoamyl alcohol | 1177 ± 1 | 52.97 ± 0.06 |
| Isobutanol | 2420 ± 20 | 108 ± 1 |
| Methanol | 128 ± 2 | 5.64 ± 0.09 |
| Triethylene glycol | 2680 ± 70 | 164 ± 4 |

**Table S2. Experimentally tested proton donors and details of experimental results.** Proton donors were tested at several concentrations (see Supplementary methods); the concentration at which the highest FE was obtained is reported along with the FE. If ammonia quantification solutions required extraction (see Supplementary methods), it is reported below. The binary activity classification is also given.

| Compound Name | Max FE (%) | Max FE error (%) | Conc. at max FE (M) | Charge (C) | Solvent used for extraction | Experimental activity classification |
|---|---|---|---|---|---|---|
| 1,2-propanediol | 0.04 | 0.02 | 0.2 | 7.2 | None | FALSE |
| 1,3-butanediol | 1.65 | 0.17 | 0.2 | 7.2 | None | TRUE |
| 1,3-propanediol | 8.38 | 1.51 | 0.1 | 7.2 | None | TRUE |
| 1,4-cyclohexane dimethanol | 0.02 | 0.05 | 0.2 | 7.2 | None | FALSE |

21

| Compound Name | Max FE (%) | Max FE error (%) | Conc. at max FE (M) | Charge (C) | Extraction solvent | Experimental activity classification |
|---|---|---|---|---|---|---|
| 1,5-pentanediol | 4.43 | 1.39 | 0.2 | 1.5 | None | TRUE |
| 1-butanol | 15.58 | 5.29 | 0.1 | 7.2 | None | TRUE |
| 1-decanol | 0.09 | 0.05 | 1 | 7.2 | Hexane | FALSE |
| 1-heptanol | 2.19 | 0.08 | 1 | 7.2 | Hexane | TRUE |
| 1-hexanol | 7.79 | 0.55 | 0.6 | 7.2 | Hexane | TRUE |
| 1-nonanol | 0.18 | 0.08 | 0.6 | 7.2 | Hexane | FALSE |
| 1-octanol | 0.08 | 0.05 | 0.2 | 7.2 | Hexane | FALSE |
| 1-pentanol | 10.42 | 3.06 | 0.2 | 7.2 | None | TRUE |
| 1-phenylethanol | 1.02 | 0.20 | 0.8 | 7.2 | None | TRUE |
| 1-propanol | 9.93 | 1.20 | 0.1 | 7.2 | None | TRUE |
| 2,2,2-trifluoroethanol | 0.02 | 0.06 | 0.4 | 7.2 | None | FALSE |
| 2,2-difluoroethanol | 0.02 | 0.00 | 0.5 | 7.2 | None | FALSE |
| 2,2-dimethyl-1,3-propanediol | 0.84 | 0.17 | 0.4 | 7.2 | None | TRUE |
| 2-butanol | 1.36 | 0.06 | 1 | 7.2 | None | TRUE |
| 2-chloroethanol | 0.06 | 0.02 | 0.2 | 7.2 | None | FALSE |
| 2-ethyl-1-butanol | 3.62 | 0.59 | 0.2 | 7.2 | None | TRUE |
| 2-phenylethanol | 1.64 | 0.26 | 0.9 | 7.2 | None | TRUE |
| 3-butene-1-ol | 1.94 | 0.08 | 0.6 | 7.2 | None | TRUE |
| 4-methoxybutan-1-ol | 0.34 | 0.02 | 0.4 | 7.2 | None | FALSE |
| Acetic acid | 0.19 | 0.07 | 0.07 | 7.2 | None | FALSE |
| Allyl alcohol | 0.69 | 0.14 | 0.6 | 7.2 | None | TRUE |
| Benzyl alcohol | 0.34 | 0.21 | 0.8 | 7.2 | None | FALSE |
| Cyclohexanol | 0.00 | 0.04 | 0.6 | 7.2 | None | FALSE |
| Ethanol | 13.16 | 1.27 | 0.1 | 7.2 | None | TRUE |
| Ethyl acetate | 0.15 | 0.06 | 0.2 | 7.2 | DCM | FALSE |
| Ethylene glycol | 0.44 | 0.03 | 0.4 | 4.1 | None | FALSE |
| Formic acid | 0.00 | 0.07 | 0.2 | 7.2 | None | FALSE |
| Glycerol | 4.20 | 0.45 | 0.2 | 4.6 | None | TRUE |
| Hexafluoro isopropyl alcohol | 0.03 | 0.09 | 0.4 | 7.2 | None | FALSE |
| Hexanoic acid | 0.00 | 0.07 | 0.2 | 7.2 | None | FALSE |
| Isoamyl alcohol | 6.74 | 2.24 | 0.6 | 7.2 | None | TRUE |
| Isobutanol | 3.09 | 0.43 | 0.4 | 7.2 | None | TRUE |
| Isopropyl alcohol | 3.12 | 0.61 | 0.2 | 7.2 | None | TRUE |
| Lactic acid | -0.03 | 0.22 | 0.1 | 7.2 | None | FALSE |

| Compound Name | Max FE (%) | Max FE error (%) | Conc. at max FE (M) | Charge (C) | Extraction solvent | Experimental activity classification |
|---|---|---|---|---|---|---|
| Methanol | 6.55 | 1.41 | 0.2 | 7.2 | None | TRUE |
| Phenol | -0.05 | 0.06 | 0.2 | 7.2 | None | FALSE |
| Propanethiol | 0.03 | 0.03 | 0.1 | 7.2 | DCM | FALSE |
| t-butyl alcohol | 0.57 | 0.08 | 0.6 | 7.2 | None | FALSE |
| Triethylene glycol | -0.15 | 0.03 | 0.2 | 7.2 | None | FALSE |
| Water | -0.06 | 0.02 | 0.2 | 7.2 | None | FALSE |

Table S3. A subset of the initial input data for the activity classification model. Here, Kamlet-Taft ($\alpha$, $\beta$, $\pi^*$) parameters obtained from experimental reports in the literature (*15–17*) are given and the corresponding experimental activity classification obtained in the current work. Additional parameters were also fed to the initial classification model to determine their important (Supplementary Fig. 1, Supplementary methods).

| Compound Name | PubChem CID | $\alpha$ | $\beta$ | $\pi^*$ | Experimental Activity Classification |
|---|---|---|---|---|---|
| 1,3-butanediol | 7966 | 0.66 | 0.84 | 0.45 | 0 |
| 1,5-pentanediol | 244 | 0.6 | 0.52 | 0.98 | 0 |
| 1-butanol | 263 | 0.84 | 0.84 | 0.47 | 1 |
| 1-heptanol | 8129 | 0.33 | 0.45 | 0.4 | 0 |
| 1-hexanol | 8103 | 0.8 | 0.84 | 0.4 | 1 |
| 1-octanol | 957 | 0.77 | 0.81 | 0.4 | 0 |
| 1-pentanol | 6276 | 0.84 | 0.86 | 0.4 | 1 |
| 1-phenylethanol | 8892 | 1.22 | 0.45 | 0.52 | 0 |
| 1-propanol | 1031 | 0.84 | 0.9 | 0.52 | 1 |
| 2-phenylethanol | 11005 | 0.55 | 0.45 | 0.36 | 0 |
| 2,2,2-trifluoroethanol | 6409 | 1.51 | 0 | 0.73 | 0 |
| Acetic acid | 176 | 1.12 | 0.45 | 0.64 | 0 |
| Allyl alcohol | 7858 | 0.84 | 0.9 | 0.52 | 0 |
| Benzyl alcohol | 996 | 1.65 | 0.3 | 0.72 | 0 |
| Benzyl alcohol | 244 | 0.6 | 0.52 | 0.98 | 0 |
| Cyclohexanol | 342 | 1.13 | 0.34 | 0.68 | 0 |
| Ethanol | 702 | 0.86 | 0.75 | 0.54 | 1 |
| Ethylene glycol | 174 | 0.9 | 0.52 | 0.92 | 0 |
| Ethylene glycol | 174 | 0.9 | 0.52 | 0.92 | 0 |
| Hexafluoro isopropyl alcohol | 13529 | 1.96 | 0 | 0.65 | 0 |
| Hexanoic acid | 8892 | 1.22 | 0.45 | 0.52 | 0 |
| Isoamyl alcohol | 31260 | 0.84 | 0.86 | 0.4 | 1 |

| Compound Name | PubChem CID | α | β | π* | Experimental Activity Classification |
|---|---|---|---|---|---|
| Isobutanol | 6560 | 0.79 | 0.84 | 0.4 | 0 |
| Isopropyl alcohol | 3776 | 0.76 | 0.84 | 0.48 | 0 |
| Lactic acid | 612 | - | - | - | 0 |
| Methanol | 887 | 0.98 | 0.66 | 0.6 | 1 |
| Phenol | 6054 | 0.64 | 0.61 | 0.88 | 0 |
| Propanethiol | 7848 | - | - | - | 0 |
| t-butyl alcohol | 6386 | 0.42 | 0.93 | 0.41 | 0 |
| Water | 962 | 1.17 | 0.47 | 1.09 | 0 |

Table S4. **The final input data for the activity classification model**. Here, Kamlet-Taft (α, β) parameters obtained from experimental reports in the literature (*15–17*) are given for a larger set of compounds and the corresponding experimental activity classification obtained in the current work. For compounds for which experimentally measured KT parameters are not known, the values predicted from the deep-learning. The set of compounds here include compounds that were initially tested and compounds which were suggested by the data-driven approach for testing. Note that experimental values for KT parameters for some compounds were not known, so predicted values from the deep learning model were given in italics for these compounds.

| Compound Name | PubChem CID | α | β | Experimental Activity Classification |
|---|---|---|---|---|
| 1,2-propanediol | 1030 | - | - | 0 |
| 1,3-butanediol | 7966 | 0.66 | 0.84 | 0 |
| 1,3-butanediol | 7896 | - | - | 1 |
| 1,3-propanediol | 10442 | - | - | 1 |
| 1,5-pentanediol | 244 | 0.6 | 0.52 | 0 |
| 1-butanol | 263 | 0.84 | 0.84 | 1 |
| 1-decanol | 8174 | 0.7 | 0.82 | 0 |
| 1-heptanol | 8129 | 0.33 | 0.45 | 0 |
| 1-hexanol | 8103 | 0.8 | 0.84 | 1 |
| 1-nonanol | 8914 | - | - | 0 |
| 1-octanol | 957 | 0.77 | 0.81 | 0 |
| 1-pentanol | 6276 | 0.84 | 0.86 | 1 |
| 1-phenylethanol | 8892 | 1.22 | 0.45 | 0 |
| 1-propanol | 1031 | 0.84 | 0.9 | 1 |
| 2-butanol | 6568 | 0.69 | 0.8 | 1 |
| 2-chloroethanol | 34 | 1.28 | 0.53 | 0 |
| 2-methoxyethanol | 8019 | - | - | 0 |
| 2-phenylethanol | 11005 | 0.55 | 0.45 | 0 |

| Compound Name | PubChem CID | α | β | Experimental Activity Classification |
|---|---|---|---|---|
| 2,2,2-trifluoroethanol | 6409 | 1.51 | 0 | 0 |
| 3-butene-1-ol | 69389 | - | - | 1 |
| Acetic acid | 176 | 1.12 | 0.45 | 0 |
| Allyl alcohol | 7858 | 0.84 | 0.9 | 0 |
| Benzyl alcohol | 996 | 1.65 | 0.3 | 0 |
| Benzyl alcohol | 244 | 0.6 | 0.52 | 0 |
| Cyclohexanol | 342 | 1.13 | 0.34 | 0 |
| Ethanol | 702 | 0.86 | 0.75 | 1 |
| Ethylene glycol | 174 | 0.9 | 0.52 | 0 |
| Ethylene glycol | 174 | 0.9 | 0.52 | 0 |
| Glycerol | 753 | 1.21 | 0.51 | 1 |
| Hexafluoro isopropyl alcohol | 13529 | 1.96 | 0 | 0 |
| Hexanoic acid | 8892 | 1.22 | 0.45 | 0 |
| Isoamyl alcohol | 31260 | 0.84 | 0.86 | 1 |
| Isobutanol | 6560 | 0.79 | 0.84 | 0 |
| Isopropyl alcohol | 3776 | 0.76 | 0.84 | 0 |
| Lactic acid | 612 | - | - | 0 |
| Methanol | 887 | 0.98 | 0.66 | 1 |
| Phenol | 6054 | 0.64 | 0.61 | 0 |
| Propanethiol | 7848 | - | - | 0 |
| t-butyl alcohol | 6386 | 0.42 | 0.93 | 0 |
| Water | 962 | 1.17 | 0.47 | 0 |

Table S5. **Input Data for the Deep Learning Prediction Model** Kamlet-Taft ($\alpha$ and $\beta$) parameters obtained from experimental reports in the literature (*15–17*).

| Compound Name | PubChem CID | α | β |
|---|---|---|---|
| 1,1,1,3,3,3-hexafluoro-2-propanol | 13529 | 1.96 | 0 |
| 1,1,1-trichloroethane | 6278 | 0 | 0 |
| 1,1,2,2-tetrachloroethane | 6591 | 0 | 0 |
| 1,1,3,3-tetramethylguanidine | 66460 | 0 | 0.86 |
| 1,1-dichloroethane | 6365 | 0.1 | 0.1 |
| 1,2,3-propanetriol | 753 | 1.21 | 0.51 |
| 1,2-diaminoethane | 3301 | 0.13 | 1.43 |
| 1,2-dibromoethane | 7839 | 0 | 0 |
| 1,2-dichlorobenzene | 7239 | 0 | 0.03 |
| 1,2-dichloroethane | 11 | 0 | 0.1 |
| 1,2-dimethoxyethane | 8071 | 0 | 0.41 |
| 1,3,5-trimethylbenzene | 7947 | 0 | 0.13 |
| 1,3-dichlorobenzene | 10943 | 0 | 0.03 |

| Compound Name | PubChem CID | α | β |
|---|---|---|---|
| 1,3-dimethylbenzene | 7929 | 0 | 0.11 |
| 1,3-dioxolane | 12586 | 0 | 0.45 |
| 1,4-difluorobenzene | 10892 | 0 | 0.03 |
| 1,4-dimethylbenzene | 7809 | 0 | 0.12 |
| 1-bromobutane | 8002 | 0 | 0.13 |
| 1-chlorobutane | 8005 | 0 | 0 |
| 1-iodobutane | 10962 | 0 | 0.23 |
| 2,2,2-trifluoroethanol | 6409 | 1.51 | 0 |
| 2,3,4-trifluoronitrobenzene | 69871 | 0 | 0.24 |
| 2,3-difluoronitrobenzene | 81335 | 0 | 0.26 |
| 2,6-dimethylpyridine | 7937 | 0 | 0.76 |
| 2-bromoacetophenone | 6259 | 0 | 0.45 |
| 2-bromopyridine | 7973 | 0 | 0.53 |
| 2-butanol | 6568 | 0.69 | 0.8 |
| 2-butanone | 6569 | 0.06 | 0.48 |
| 2-chloroacetophenone | 10757 | 0 | 0.45 |
| 2-chloroaniline | 7240 | 0.25 | 0.4 |
| 2-chlorobenzaldehyde | 6996 | 0 | 0.4 |
| 2-chloroethanol | 34 | 1.28 | 0.53 |
| 2-cyanopyridine | 7522 | 0 | 0.29 |
| 2-decanone | 12741 | 0 | 0.48 |
| 2-fluoroacetophenone | 96744 | 0 | 0.47 |
| 2-fluoronitrobenzene | 73895 | 0 | 0.28 |
| 2-fluoropyridine | 9746 | 0 | 0.51 |
| 2-heptanone | 8051 | 0.05 | 0.48 |
| 2-methyl-1-propanol | 6560 | 0.79 | 0.84 |
| 2-methyl-2-butanol | 6405 | 0.28 | 0.93 |
| 2-methyl-2-propanol | 6386 | 0.42 | 0.93 |
| 2-methylnitrobenzene | 6944 | 0 | 0.3 |
| 2-nonanone | 13187 | 0 | 0.48 |
| 2-octanone | 8093 | 0 | 0.48 |
| 2-pentanone | 7895 | 0 | 0.52 |
| 2-phenylacetonitrile | 8794 | 0 | 0.41 |
| 2-phenylethanol | 6054 | 0.64 | 0.61 |
| 2-propanol | 3776 | 0.76 | 0.84 |
| 2-propanone | 180 | 0.08 | 0.43 |
| 2-propen-1-ol | 7858 | 0.84 | 0.9 |
| 2-pyrrolidone | 12025 | 0.36 | 0.77 |
| 3,4,5-trifluoronitrobenzene | 2782793 | 0 | 0.24 |
| 3,4-difluoronitrobenzene | 123053 | 0 | 0.26 |

| Compound Name | PubChem CID | α | β |
|---|---|---|---|
| 3,4-dimethylpyridine | 11417 | 0 | 0.78 |
| 3-bromoacetophenone | 16502 | 0 | 0.45 |
| 3-bromopyridine | 12286 | 0 | 0.6 |
| 3-chlorobenzaldehyde | 11477 | 0 | 0.4 |
| 3-chlorophenol | 7933 | 1.57 | 0.23 |
| 3-fluoroacetophenone | 9967 | 0 | 0.47 |
| 3-fluoronitrobenzene | 9823 | 0 | 0.28 |
| 3-methyl-1-butanol | 31260 | 0.84 | 0.86 |
| 3-methylphenol | 342 | 1.13 | 0.34 |
| 3-pentanone | 7288 | 0 | 0.45 |
| 3-phenylpropanol | 31234 | 0.53 | 0.55 |
| 3-trifluoromethylnitrobenzene | 7386 | 0 | 0.25 |
| 4-chlorobenzaldehyde | 7726 | 0 | 0.4 |
| 4-fluoronitrobenzene | 9590 | 0 | 0.28 |
| 4-methyl-2-oxo-1,3-dioxolane | 7924 | 0 | 0.4 |
| 4-methyl-2-pentanone | 7909 | 0.02 | 0.48 |
| 4-methylphenol | 2879 | 1.64 | 0.34 |
| 4-methylpyridine | 7963 | 0 | 0.67 |
| acetic acid | 176 | 1.12 | 0.45 |
| acetic anhydride | 7918 | 0 | 0.29 |
| acetonitrile | 6342 | 0.19 | 0.4 |
| acetophenone | 7410 | 0.04 | 0.49 |
| aniline | 6115 | 0.26 | 0.5 |
| benzene | 241 | 0 | 0.1 |
| benzonitrile | 7505 | 0 | 0.37 |
| benzyl alcohol | 244 | 0.6 | 0.52 |
| bis(2-chloroethyl) ether | 8115 | 0 | 0.4 |
| bis(2-methoxyethyl) ether | 8150 | 0 | 0.4 |
| bromobenzene | 7961 | 0 | 0.06 |
| butane | 7843 | 0 | 0 |
| butanenitrile | 8008 | 0 | 0.4 |
| butanoic acid | 264 | 1.1 | 0.45 |
| butanol | 263 | 0.84 | 0.84 |
| butyl acetate | 31272 | 0 | 0.45 |
| butylamine | 8007 | 0 | 0.72 |
| carbon disulfide | 6348 | 0 | 0.07 |
| chloroacetonitrile | 7856 | 0 | 0.34 |
| chlorobenzene | 7964 | 0 | 0.07 |
| cis-decalin | 7044 | 0 | 0.08 |
| cyclohexane | 8078 | 0 | 0 |

| Compound Name | PubChem CID | α | β |
|---|---|---|---|
| cyclohexanol | 7966 | 0.66 | 0.84 |
| cyclohexanone | 7967 | 0 | 0.53 |
| cyclopentanone | 8452 | 0 | 0.52 |
| decane | 15600 | 0 | 0 |
| decanol | 8174 | 0.7 | 0.82 |
| diaminoethane | 3301 | 0.13 | 1.43 |
| dibenzyl ether | 7657 | 0 | 0.41 |
| dibromomethane | 3024 | 0 | 0 |
| dibutyl ether | 8909 | 0 | 0.46 |
| dichloromethane | 6344 | 0.13 | 0.1 |
| diethyl carbonate | 7766 | 0 | 0.4 |
| diethyl ether | 3283 | 0 | 0.47 |
| diethyl sulfide | 9609 | 0 | 0.37 |
| diethylamine | 8021 | 0.3 | 0.7 |
| diethylformamide | 12051 | 0 | 0.79 |
| diiodomethane | 6346 | 0 | 0 |
| diisopropyl ether | 7914 | 0 | 0.49 |
| diisopropyl sulfide | 12264 | 0 | 0.38 |
| dimethyl carbonate | 12021 | 0 | 0.43 |
| dimethyl phthalate | 8554 | 0 | 0.78 |
| dimethyl sulfate | 6497 | 0 | 0.36 |
| dimethyl sulfide | 1068 | 0 | 0.34 |
| dimethyl sulfoxide | 679 | 0 | 0.76 |
| dimethylamine | 674 | 0 | 0.7 |
| dimethylcyanamide | 15112 | 0 | 0.64 |
| di-n-butyl sulfide | 11002 | 0 | 0.38 |
| dioxane | 31275 | 0 | 0.37 |
| dipentylamine | 16316 | 0 | 0.7 |
| dipentylether | 12743 | 0 | 0.47 |
| diphenyl ether | 7583 | 0 | 0.13 |
| dipropyl ether | 8114 | 0 | 0.46 |
| dodecane | 8182 | 0 | 0 |
| ethane | 6324 | 0 | 0 |
| ethanediol | 174 | 0.9 | 0.52 |
| ethanol | 702 | 0.86 | 0.75 |
| ethoxybenzene | 7674 | 0 | 0.3 |
| ethyl acetate | 8857 | 0 | 0.45 |
| ethyl benzoate | 7165 | 0 | 0.41 |
| ethyl chloroacetate | 7751 | 0 | 0.35 |
| ethyl formate | 8025 | 0 | 0.36 |

| Compound Name | PubChem CID | α | β |
|---|---|---|---|
| ethyl trichloroacetate | 10588 | 0 | 0.25 |
| ethylene carbonate | 7303 | 0 | 0.41 |
| fluorobenzene | 10008 | 0 | 0.07 |
| formamide | 713 | 0.71 | 0.48 |
| formic acid | 284 | 1.23 | 0.38 |
| furan | 8029 | 0 | 0.14 |
| heptane | 8900 | 0 | 0 |
| heptanoic acid | 8094 | 1.2 | 0.45 |
| hexachlorobiphenyl | 91635 | 0 | 0.03 |
| hexadecane | 11006 | 0 | 0 |
| hexafluorobenzene | 9805 | 0 | 0.02 |
| hexamethylphosphoramide | 12679 | 0 | 1.05 |
| hexane | 8058 | 0 | 0 |
| hexanoic acid | 8892 | 1.22 | 0.45 |
| hexanol | 8103 | 0.8 | 0.84 |
| iodobenzene | 11575 | 0 | 0.06 |
| methane | 297 | 0 | 0 |
| methanol | 887 | 0.98 | 0.66 |
| methoxybenzene | 7519 | 0 | 0.32 |
| methyl acetate | 6584 | 0 | 0.42 |
| methyl benzoate | 7150 | 0 | 0.38 |
| methyl formate | 7865 | 0 | 0.37 |
| methyl propanoate | 11124 | 0 | 0.27 |
| methylamine | 6329 | 0 | 0.7 |
| morpholine | 8083 | 0.29 | 0.7 |
| N,N,N',N'-tetramethylurea | 12437 | 0 | 0.8 |
| N,N-diethylacetamide | 12703 | 0 | 0.78 |
| N,N-dimethylacetamide | 31374 | 0 | 0.76 |
| N,N-dimethylaniline | 949 | 0 | 0.43 |
| N,N-dimethylbenzylamine | 7681 | 0 | 0.64 |
| N,N-dimethylcyclohexylamine | 7415 | 0 | 0.84 |
| N,N-dimethylformamide | 6228 | 0 | 0.69 |
| n-decylamine | 8916 | 0 | 0.68 |
| n-heptylamine | 8127 | 0 | 0.69 |
| nitrobenzene | 7416 | 0 | 0.3 |
| nitromethane | 6375 | 0.22 | 0.06 |
| N-methylacetamide | 6582 | 0.47 | 0.8 |
| N-methylaniline | 7515 | 0.17 | 0.47 |
| N-methylformamide | 31254 | 0.62 | 0.8 |
| N-methylpyrrolidon | 13387 | 0 | 0.77 |

| Compound Name | PubChem CID | α | β |
|---|---|---|---|
| n-nonylamine | 16215 | 0 | 0.69 |
| n-octylamine | 8143 | 0 | 0.69 |
| nonane | 8141 | 0 | 0 |
| octadecafluoronaphthalene | 9386 | 0 | -0.05 |
| octane | 356 | 0 | 0 |
| octanol | 957 | 0.77 | 0.81 |
| oxane | 8894 | 0 | 0.54 |
| oxolan-2-one | 7302 | 0 | 0.49 |
| pentachlorobiphenyl | 17348 | 0 | 0.06 |
| pentadecane | 12391 | 0 | 0 |
| pentafluoropyridine | 69690 | 0 | 0.16 |
| pentane | 8003 | 0 | 0 |
| pentanoic acid | 7991 | 1.19 | 0.45 |
| pentanol | 6276 | 0.84 | 0.86 |
| perfluoro(methylcyclohexane) | 9637 | 0 | -0.06 |
| phenol | 996 | 1.65 | 0.3 |
| piperidine | 8082 | 0 | 1.04 |
| propane | 6334 | 0 | 0 |
| propanenitrile | 7854 | 0 | 0.39 |
| propanoic acid | 1032 | 1.12 | 0.45 |
| propanol | 1031 | 0.84 | 0.9 |
| propyl acetate | 7997 | 0 | 0.4 |
| pyridine | 1049 | 0 | 0.64 |
| pyrrolidine | 31268 | 0.16 | 0.7 |
| quinoline | 7047 | 0 | 0.64 |
| styrene | 7501 | 0 | 0.12 |
| sulfolane | 31347 | 0 | 0.39 |
| tetrachloroethene | 31373 | 0 | 0.05 |
| tetrachloromethane | 5943 | 0 | 0.1 |
| tetradecafluorohexane | 9639 | 0 | -0.08 |
| tetradecane | 12389 | 0 | 0 |
| tetrahydrofuran | 8028 | 0 | 0.55 |
| tetramethylsilane | 6396 | 0 | 0.02 |
| thiane | 15367 | 0 | 0.36 |
| thiolane | 1127 | 0 | 0.44 |
| thiolane-1-oxide | 1128 | 0 | 0.81 |
| toluene | 1140 | 0 | 0.11 |
| trans-1,2-dichloroethene | 638186 | 0 | 0 |
| tribromomethane | 5558 | 0.05 | 0.05 |
| tributyl phosphate | 31357 | 0 | 0.8 |

| Compound Name | PubChem CID | α | β |
|---|---|---|---|
| tributylamine | 7622 | 0 | 0.62 |
| trichloroethene | 6575 | 0 | 0.05 |
| trichloromethane | 6212 | 0.2 | 0.1 |
| tridecane | 12388 | 0 | 0 |
| triethyl phosphate | 6535 | 0 | 0.77 |
| triethylamine | 8471 | 0 | 0.71 |
| trimethyl phosphate | 10541 | 0 | 0.77 |
| undecane | 14257 | 0 | 0 |
| water | 962 | 1.17 | 0.47 |

# References

1. N. Lazouski, Z. J. Schiffer, K. Williams, K. Manthiram, *Joule* **3**, 1127 (2019).

2. J. Quinlan, *San Mateo* (1993).

3. L. Breiman, J. Friedman, C. J. Stone, R. A. Olshen, *Classification and regression trees* (CRC press, 1984).

4. G. V. Kass, *Journal of the Royal Statistical Society: Series C (Applied Statistics)* **29**, 119 (1980).

5. H. Kim, W.-Y. Loh, *Journal of the American Statistical Association* **96**, 589 (2001).

6. H. Kim, W.-Y. Loh, *Journal of Computational and Graphical Statistics* **12**, 512 (2003).

7. W.-Y. Loh, *The Annals of Applied Statistics* pp. 1710–1737 (2009).

8. W.-Y. Loh, Y.-S. Shih, *Statistica sinica* pp. 815–840 (1997).

9. A. Fielding, C. O'Muircheartaigh, *Journal of the Royal Statistical Society: Series D (The Statistician)* **26**, 17 (1977).

10. R. Messenger, L. Mandell, *Journal of the American statistical association* **67**, 768 (1972).

11. W.-Y. Loh, *Wiley Interdisciplinary Reviews: Data Mining and Knowledge Discovery* **1**, 14 (2011).

12. R. Timofeev, *Humboldt University, Berlin* pp. 1–40 (2004).

13. S. Kearnes, K. McCloskey, M. Berndl, V. Pande, P. Riley, *Journal of computer-aided molecular design* **30**, 595 (2016).

14. Z. Wu, *et al.*, *Chem. Sci.* **9**, 513 (2018).

15. Y. Marcus, *Chemical Society Reviews* **22**, 409 (1993).

16. P. Meyer, G. Maurer, *Industrial & engineering chemistry research* **34**, 373 (1995).

17. R. Stenutz, Kamlet-taft solvent parameters, `http://www.stenutz.eu/chem/solv26.php` (accessed June 2020).



\end{document}